\documentclass[10pt,conference]{IEEEtran}
\IEEEoverridecommandlockouts

\usepackage{cite}
\usepackage{booktabs}

\usepackage{amsmath,amssymb,amsfonts,amsthm}

\usepackage[ruled,linesnumbered]{algorithm2e}
\SetKwComment{Comment}{/* }{ */}

\usepackage{fancyhdr}
\theoremstyle{definition}
\newtheorem{definition}{Def.}
\newtheorem{observation}{Obs.}

\usepackage{caption}
\usepackage{graphicx}
\usepackage{textcomp}
\usepackage{xcolor}
\usepackage{tikz}

\newcommand*\circled[1]{\tikz[baseline=(char.base)]{
\node[shape=circle,draw,inner sep=1pt] (char) {#1};}}
\def\BibTeX{{\rm B\kern-.05em{\sc i\kern-.025em b}\kern-.08em
T\kern-.1667em\lower.7ex\hbox{E}\kern-.125emX}}

\usepackage{hyperref}
\hypersetup{
    colorlinks=true,
    linkcolor=blue,
    filecolor=magenta,      
    urlcolor=cyan,
    }
\usepackage{multicol}
\usepackage{multirow}

\usepackage{listings}

\lstdefinestyle{mystyle}{
    backgroundcolor=\color{backcolour},   
    commentstyle=\color{codegreen},
    keywordstyle=\color{magenta},
    numberstyle=\tiny\color{codegray},
    stringstyle=\color{codepurple},
    basicstyle=\ttfamily\footnotesize,
    breakatwhitespace=false,         
    breaklines=true,                 
    captionpos=b,                    
    keepspaces=true,                 
    numbers=left,                    
    numbersep=5pt,                  
    showspaces=false,                
    showstringspaces=false,
    showtabs=false,                  
    tabsize=2
}
\usepackage{enumitem}
\setlist[itemize]{align=parleft,left=0pt..1em}
\usepackage[normalem]{ulem}
\usepackage{framed}
\definecolor{shadecolor}{gray}{0.9}

\newcommand*{\affaddr}[1]{#1} 

\newcommand*{\email}[1]{\texttt{#1}}

\begin{document}

\title{BOXR: Body and head motion Optimization framework for eXtended Reality}

\author{%
Ziliang Zhang, Zexin Li, Hyoseung Kim, Cong Liu\\
\affaddr{University of California, Riverside}\\
\email{\{zzhan357, zli536, hyoseung, congl\}@ucr.edu}%
}
 
\maketitle

\thispagestyle{fancy}

\begin{abstract} 

The emergence of standalone Extended Reality (XR) systems has enhanced user mobility, accommodating both subtle, frequent head motions and substantial, less frequent body motions. However, the pervasively used Motion-to-Display (M2D) latency metric, which measures the delay between the most recent motion and its corresponding display update, only accounts for head motions. This oversight can leave users prone to motion sickness if significant body motion is involved. Although existing methods optimize M2D latency through asynchronous task scheduling and reprojection methods, they introduce challenges like resource contention between tasks and outdated pose data. These challenges are further complicated by user motion dynamics and scene changes during runtime.

To address these issues, we for the first time introduce the Camera-to-Display (C2D) latency metric, which captures the delay caused by body motions, and present BOXR, a framework designed to co-optimize both body and head motion delays within an XR system. BOXR enhances the coordination between M2D and C2D latencies by efficiently scheduling tasks to avoid contentions while maintaining an up-to-date pose in the output frame. Moreover, BOXR incorporates a motion-driven visual inertial odometer to adjust to user motion dynamics and employs scene-dependent foveated rendering to manage changes in the scene effectively.
Our evaluations show that BOXR significantly outperforms state-of-the-art solutions in 11 EuRoC MAV datasets across 4 XR applications across 3 hardware platforms. In controlled motion and scene settings, BOXR reduces M2D and C2D latencies by up to 63\% and 27\%, respectively and increases frame rate by up to 43\%. 
In practical deployments, BOXR achieves substantial reductions in real-world scenarios—up to 42\% in M2D latency and 31\% in C2D latency—while maintaining remarkably low miss rates of only 1.6\% for M2D requirements and 1.0\% for C2D requirements.

\end{abstract}
 
\section{Introduction}
Extended Reality (XR), which encompasses virtual reality, augmented reality, and mixed reality, offers an immersive virtual-physical experience to the user~\cite{openxr,apple_vision_pro}. An XR system captures the user's motion through various sensors such as an inertial measurement unit (IMU) and a camera (CAM), processes the sensor data to generate a new image frame according to the user's motion and surrounding scene changes, and displays the generated frame on a head-mounted display.

\begin{figure}[t]
\includegraphics[width=\linewidth]{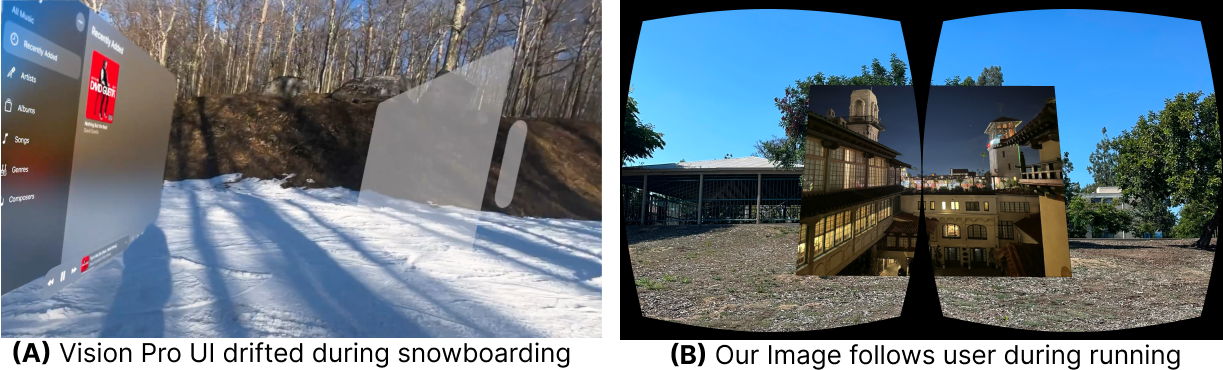}
\caption{Comparison between the Vision Pro~\cite{vision_pro_snowboarding} and BOXR when large body motion exists}
\label{fig_init_compare}
\end{figure}

As evident by prior research, the delay between a motion's capture and the motion's display in the output frame can cause perception misalignment~\cite{m2d_20_1, motion_sickness}. Existing systems address this by focusing on optimizing the Motion-to-Display latency (M2D) metric, which measures the time it takes to display the latest motion. Despite extensive optimization of M2D, as illustrated in the left subfigure of Fig.~\ref{fig_init_compare}, the virtual user interface of Apple Vision Pro~\cite{apple_vision_pro} drifts towards the opposite direction of movement when the user makes significant movements during snowboarding~\cite{vision_pro_snowboarding}. This significantly hinders the quality of the immersive experience and can cause motion sickness.

We find that the fundamental reason behind this problem lies in the imprecise distinction of ``motion''. The user's motion recognized by an XR system can be categorized into two types based on the required sensors. The first type of motion includes frequent rotational movements or small-magnitude translational movements that do not involve position changes within the environment and thus can be reliably captured by IMU. Since this motion primarily involves head movement, we refer to it as \textit{head motion}.
The second type of motion involves less frequent but more extensive displacement, specifically movements that require camera-based localization in the environment to capture absolute position changes, such as walking in a room. Because this motion results in changes in the user's body posture, we refer to it as \textit{body motion}.
The M2D metric widely used in current studies represents {\em only the head motion} captured by the IMU sensor running at a high sampling rate (e.g., 500 Hz). Consequently, body motion, which requires slower camera updates (e.g., 20 Hz), remains unaccounted for in M2D assessments. Therefore, even with significant optimization of M2D by Vision Pro~\cite{vision_pro_m2d} and state-of-the-art research~\cite{min_m2d,atw,hou2019head}, the interface continues to encounter drift caused by the substantial delay of body motion in the output frame.


\begin{figure*}[t]
\includegraphics[width=\textwidth]{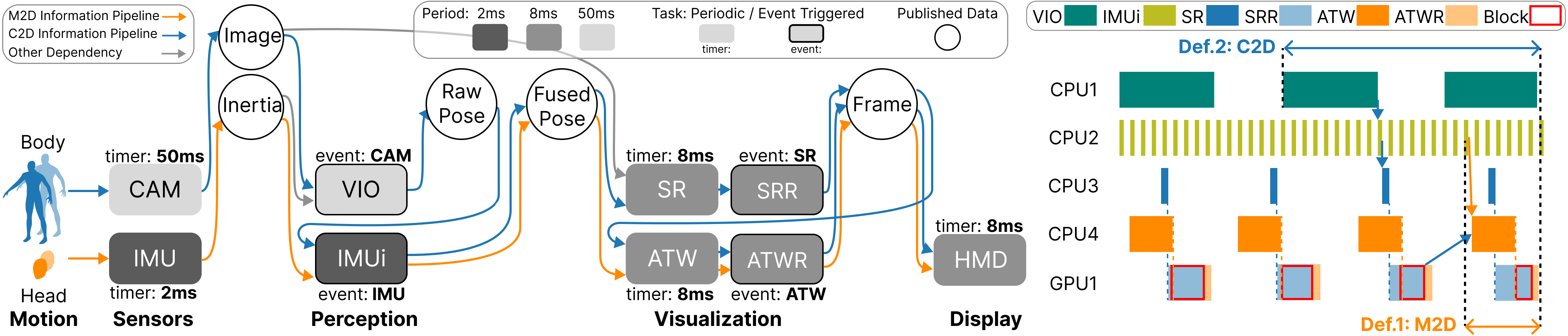}
\caption{
Left: ILLIXR, an open-sourced state-of-the-art XR system, involves two types of motion processing sequences for the final display. 
Right: Example schedule of XR tasks. Blue arrows show C2D and orange arrows show M2D sequences.
}
\label{fig_pub_sub}
\end{figure*}


After a thorough analysis of the end-to-end XR execution pipeline, we identified the root cause of why the M2D metric alone cannot capture the delay between body motion detection and its display. To address this, we define and introduce another key metric: Camera-to-Display latency (C2D) which is essential for accurately capturing this delay.
However, co-optimizing M2D and C2D in an XR system involves several challenges. 
First, state-of-the-art XR frameworks like ILLIXR~\cite{illixr} schedule tasks asynchronously and use reprojection-based methods such as asynchronous timewarp~\cite{atw,min_m2d} to optimize M2D. However, there is contention between the reprojection task required by M2D and the scene rendering task essential to C2D, as both need to compete for the shared GPU.
Straightforward solutions include adjusting the periodicity of tasks to simultaneously improve M2D and C2D cannot eliminate contention, resulting in greater fluctuations in both metrics and exacerbating motion sickness. Additionally, the dynamic nature of user motion and scene changes introduces significant variations in task execution times, further complicating this issue.


\textbf{Contributions.} To effectively co-optimize M2D and C2D, we present \textbf{BOXR}, a \textbf{B}ody and head motion delay \textbf{O}ptimization framework for e\textbf{X}tended \textbf{R}eality systems. BOXR incorporates our discovery of C2D based on the delay of body motion and addresses the unique challenges due to co-optimization. 
First, BOXR proposes a contention-preventive scheduling policy to eliminate contention between rendering and reprojection tasks. Additionally, it designs an on-demand IMUi to minimize IMU wasted work during execution. This approach ensures consistently low M2D and C2D latencies, through efficiently managing task execution sequences to prevent unnecessary delays and conflicts.
Built upon the scheduling policy, BOXR designs a {\em motion-driven} visual inertial odometer (VIO) which dynamically adapts the feature extraction frontend to the motion dynamics, correcting additional position errors with an error-bounding method. 
The intriguing aspect of this design is how it intelligently adapts in real-time to the user's motions, maintaining VIO execution times within the set budget without significantly increasing position errors.
Additionally, BOXR implements Scene-Dependent Foveated Rendering, a technique that dynamically adjusts the foveation area according to the number of objects in the scene. By centering foveation on the centroid of objects, BOXR manages to maintain high frame quality while controlling render times. Our intuition is to prioritize rendering resources where they matter most, based on real-time scene analysis.

We implement BOXR based on the ILLIXR framework~\cite{illixr} and test four popular XR applications across three hardware settings with eleven trajectories\footnote{Our implementation can be found at: https://github.com/rtenlab/BOXR.git}. As shown in Fig.~\ref{fig_init_compare}, BOXR consistently maintains the interface at the center of the viewport that closely aligns with user movement when significant body motion occurs. We conduct evaluations using both controlled trajectory datasets with diverse XR application scenes and real-world scenarios featuring varied motion levels. Compared to the state-of-the-art, BOXR achieves:
\begin{itemize}
    \item \textbf{Effectiveness:} BOXR reduces up to 63\% of M2D and 27\% of C2D while increasing frame rate by up to 43\% across three hardware platforms under different applications and trajectories.
    \item \textbf{Robustness:} In extremely dynamic motion and scene scenarios, BOXR experiences at most 10\% M2D increase and 6\% C2D increase, leading to stable usability.
    \item \textbf{Applicability:} When deploying in real-world scenarios, BOXR achieves up to 42\% M2D reduction and 31\% C2D reduction, resulting in only 1.6\% M2D miss rate and 1.0\% C2D miss rate across three applications.
\end{itemize}

\section{Background}
\subsection{XR System Model}
\label{subsec_xr_system_model}


Recent XR systems involve various sensing and algorithmic tasks. In this subsection, we give a brief description of these tasks and their execution workflow in ILLIXR~\cite{illixr}, which is a representative state-of-the-art open-source XR framework. Fig. \ref{fig_pub_sub} illustrates the overview of the ILLIXR system.

To maximize utilization and increase throughput, ILLIXR adopts a publisher-subscriber model, where each task runs asynchronously as shown in Fig.~\ref{fig_pub_sub} right~\cite{illixr,min_m2d}. Sensor readings from the IMU and CAM are published onto image and inertia topics and pipelined into the {\em perception phase} consisting of VIO and IMUi tasks:
\begin{itemize}
\item Visual Inertial Odometer (VIO): The VIO task subscribes to both CAM image and IMU inertia topics, but executes only when the CAM topic arrives, consequently getting the same period as CAM. VIO extracts features and conducts pose estimation based on Multi-State Constraint Kalman Filter (MSCKF) algorithm~\cite{openvins}. VIO publishes as output the estimated pose capturing the latest body movement. 
\item IMU integration (IMUi): The IMUi task is triggered whenever a new IMU sensor reading arrives~\cite{gtsam}. IMUi extrapolates the last pose from VIO to the arrival time of the IMU data, in order to compensate for the slow update rate of CAM and the long processing time of VIO. 
We call the output of IMUi a fused pose since it captures body motion from the raw pose as well as head motion using the latest inertia data from IMU.
\end{itemize}
Next, the {\em visualization phase} involves the following tasks:
\begin{itemize}
    \item Scene Reconstruction (SR): The SR task employs the fused pose from the perception phase and calculates the viewport that determines the display area at the default 120 FPS target frame rate, which is equivalent to an 8 ms period.
    \item Scene Reconstruction Rendering (SRR): The completion of the SR task activates the SRR task. SRR renders a 2D frame within the viewport provided by SR.
    \item Asynchronous Timewarp (ATW): The ATW task leverages the up-to-date fused pose published after SRR to generate the user's view matrix, which is the inverse of the camera transformation matrix that contains the complete 3D information. Determined by the default target frame rate, ATW is also scheduled with the 8 ms period and runs asynchronously to SR and SRR.
    \item Asynchronous Timewarp Reprojection (ATWR): Upon completion of ATW, the ATWR task reprojects the 2D frame based on the view matrix from ATW and completes the final 3D frame.
\end{itemize}

During the execution, VIO, IMUi, SR, and ATW only use the CPU for computation, whereas SRR and ATWR primarily use the GPU for rendering and projection. 
Each of these tasks is a separate thread and they run concurrently unless there is any explicit input dependency explained above.
In particular, SRR and ATWR are designed to share a single GPU stream because they access the same memory region of the frame buffer; hence, they are executed sequentially on the GPU in a first-in-first-out order~\cite{kato2011timegraph,wang2021balancing}. 

\subsection{Characterizing latency metrics}
\label{subsec_characterizing_latency_metrics}
Since humans are highly perceptive of motion delay, any delay between the 
latest user motion and the displayed output
that exceeds 20 ms leads to perception misalignment in an XR system~\cite{m2d_20_1,motion_sickness}. Within the context of XR, the latest `head' motion is consistently captured by IMU due to its high sample rate compared to CAM. In the existing work~\cite{atw,illixr,min_m2d,m2d_20_1}, this delay is quantified using the following metric:
\begin{definition}[Motion-to-Display Latency]
The motion-to-display latency (M2D) is defined as the time interval between capturing of the latest motion by the IMU and its corresponding display on the HMD. 
In other words, M2D is the time to complete the IMUi$\rightarrow$ATW$\rightarrow$ATWR sequence, e.g., orange arrows in both sub-diagrams of Fig.~\ref{fig_pub_sub}.
\label{def_m2d}
\end{definition}

However, as M2D does not distinguish between different types of motion, it always begins from the most recent IMU sensor input that captures only head motion. Consequently, even if the state-of-the-art XR system achieves satisfactory M2D performance~\cite{apple_vision_pro,vision_pro_m2d}, users still experience noticeable perception misalignment when engaging in significant body motion, as we showed with Fig.~\ref{fig_init_compare}(A). 
To better address the perception misalignment and effectively differentiate between delays caused by IMU-captured head motion and CAM-captured body motion, we introduce the following: 
\begin{definition}[Camera-to-Display Latency]
The camera-to-display latency (C2D) is defined as the time interval between capturing the latest body motion by CAM and its corresponding display on the HMD.\footnote{It is worth noting that C2D differs from the photon-to-photon latency~\cite{photon-to-display} used in some literature since it only measures the time for see-through display of CAM images on the HDM, without involving the pose estimation of VIO.} C2D is therefore the time to complete the sequence of VIO$\rightarrow$IMUi$\rightarrow$SR$\rightarrow$SRR$\rightarrow$ATW$\rightarrow$ATWR, e.g., blue arrows in both sub-diagrams of Fig.~\ref{fig_pub_sub}.
\label{def_c2d}
\end{definition}

C2D begins with VIO which generates a raw pose. IMUi then calculates to produce a fused pose that encompasses both head and body motion. Subsequently, SR and SRR utilize this fused pose to render the 2D frame, which ATW and ATWR further reproject to produce the final 3D frame output. 
Given that users typically perceive displacement through changes of objects in the scene, we set the C2D requirement to 80 ms, a threshold commonly applied in video games to ensure steady updates of virtual objects~\cite{openxr,c2d_60_2}.

\section{Challenges}
\label{sec_challenges}

\subsection{Challenges for M2D and C2D optimization}\label{subsec_optimization_space}

To understand the correlation between M2D and C2D, we will discuss notable performance bottlenecks and 
why simple methods that are seemingly positive in reducing latencies do not work in XR systems.
We conducted experiments using two prevalent XR applications: Gldemo~\cite{illixr}, which has lighter rendering demands, and Sponza~\cite{godot}, known for its intensive rendering requirements. Both applications are implemented on ILLIXR~\cite{illixr}. Only a PC platform is used in this section but our evaluation in Sec.~\ref{sec:eval} includes two embedded platforms as well. We used the default target frame rate of 120 FPS set by ILLIXR, which gives the task periods shown in Fig.~\ref{fig_pub_sub}.

We first analyze the effect of ATW and ATWR tasks. In existing XR systems, ATW and ATRW are known to be effective in reducing M2D, thanks to their asynchronous execution to SR and SRR~\cite{atw, illixr, min_m2d}. The right side of Fig.~\ref{fig_pub_sub} gives an example schedule. 
The execution time of SR is much smaller than that of ATW because SR calculates only one viewport position but ATW performs the entire matrix transformation based on the latest fused pose. Conversely, SRR takes longer to execute than ATWR because SRR renders a whole 2D frame based on the viewport from SR but ATWR only reprojects the scene with the depth information from ATW. Also, we can observe from the figure that the execution of SRR and ATRW does not overlap (highlighted by red blocks). This is because they share the frame buffer memory, requiring them to contend for the same GPU stream and execute sequentially.


\begin{figure}[t]
\includegraphics[width=\linewidth]{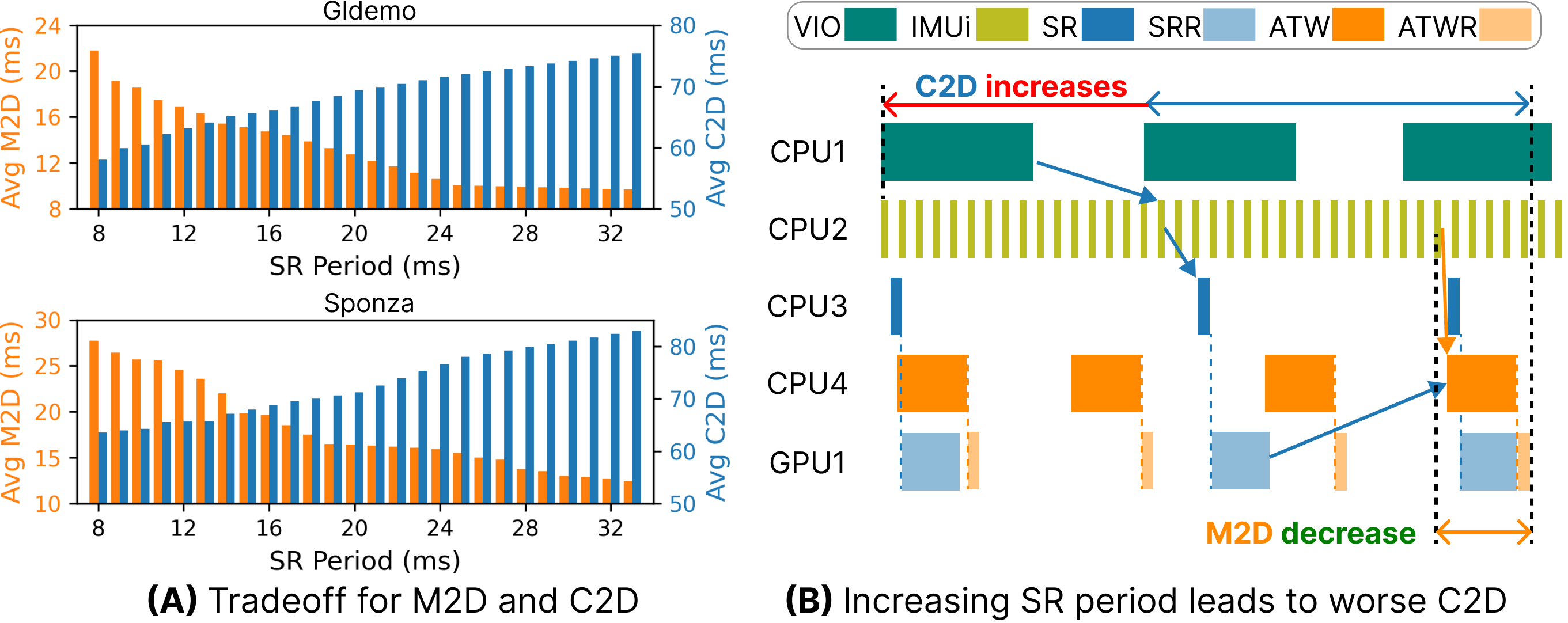}
\caption{Negative impact of improving M2D on C2D}
\label{fig_focus_m2d}
\end{figure}

To mitigate such GPU contention for ATWR and improve M2D, one may consider increasing the period of SR, thereby decreasing the number of SRR jobs and reducing the blocking time caused by contention. 
However, Fig.~\ref{fig_focus_m2d}(A) demonstrates that optimizing M2D by increasing the SR period leads to a larger C2D. 
This phenomenon is further elucidated in Fig.~\ref{fig_focus_m2d}(B). As the SR period is increased, the last ATWR job receives the raw pose from the first VIO job, which is outdated since the second VIO has already finished execution. The use of this outdated raw pose causes a significant C2D increase, the amount of which is illustrated by the red arrow in Fig.~\ref{fig_focus_m2d}(B).


\begin{observation}
Solely optimizing M2D latency by minimizing contention may lead to the deterioration of C2D because of the outdated raw pose caused by reduced SR workload. This necessitates the consideration of both M2D and C2D during optimization.
\label{obs_1}
\end{observation}


\begin{figure}[t]
\includegraphics[width=\linewidth]{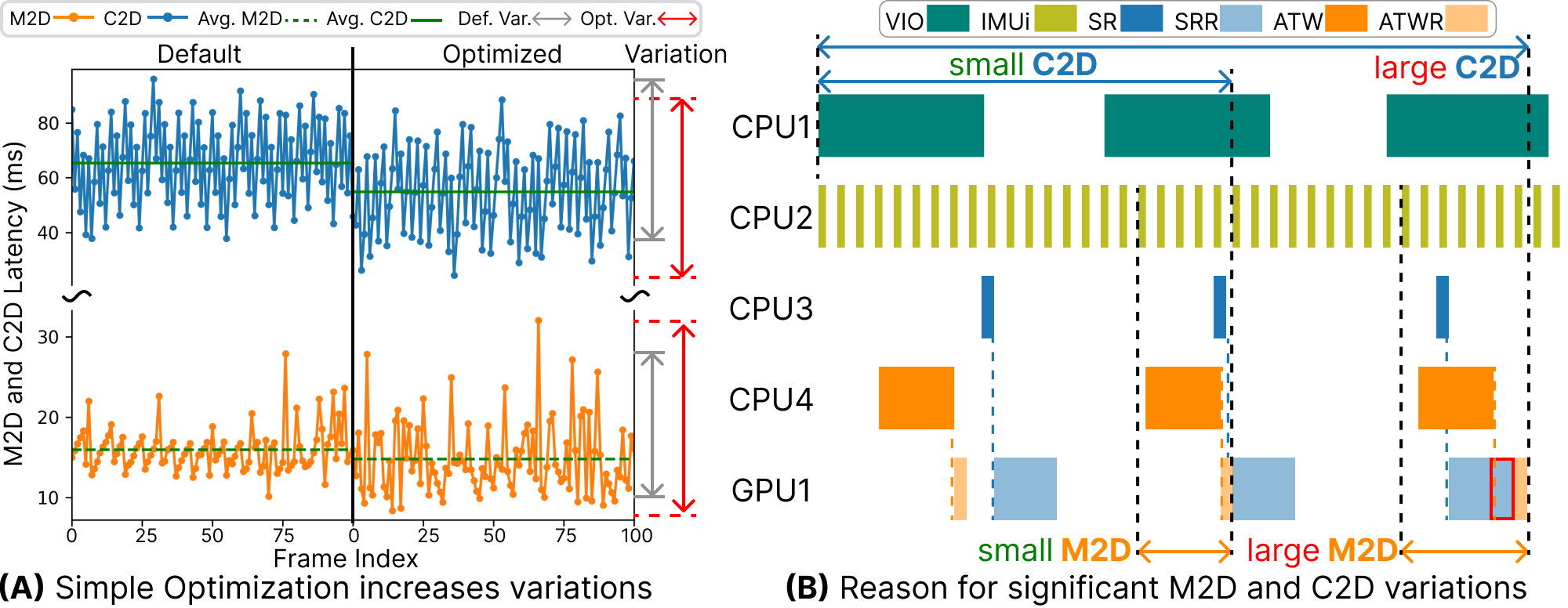}
\caption{Effects of simple co-optimization using task timers}
\label{fig_co_op_fail}
\end{figure}

A better approach might be adjusting the periods of both SR and ATW. 
However, such an approach could have side effects as detailed in Fig.~\ref{fig_co_op_fail}(A). It shows the profiling of M2D and C2D in 100 frames output when running Gldemo with default scheduling and with the optimized scheduling that achieved the smallest average-case M2D and C2D among various combinations of SR and ATR periods.
Although the optimization can achieve a lower M2D and C2D on average than the default setting, their variations are significantly increased. 
Compare Fig.~\ref{fig_co_op_fail}(B) to the right side of Fig.~\ref{fig_pub_sub}. While changing both SR and ATW periods fixes the GPU contention and the outdated raw pose issues for the first and second ATWR jobs, the third ATWR job still suffers from the same issues.
This leads to an extremely large C2D that even surpasses the M2D and C2D in Fig.~\ref{fig_focus_m2d}(B). 
Such significant fluctuations of M2D and C2D can cause great discrepancies between the actual and perceived motions, which is the root cause of motion sickness~\cite{motion_sickness}. 


\begin{observation}
Simply adjusting the periodicities of XR tasks can cause significant fluctuations in M2D and C2D. A scheduling technique that considers the freshness of sensor data and GPU contention is necessary.
\label{obs_2}
\end{observation}

\subsection{XR characteristics at runtime}

In real-world scenarios, user motion dynamics and scene diversities contribute to significant variations in the execution time of certain tasks. We identify two characteristics (C1, C2) unique to XR systems given the runtime dynamics.

\begin{figure}[t]
\includegraphics[width=\linewidth]{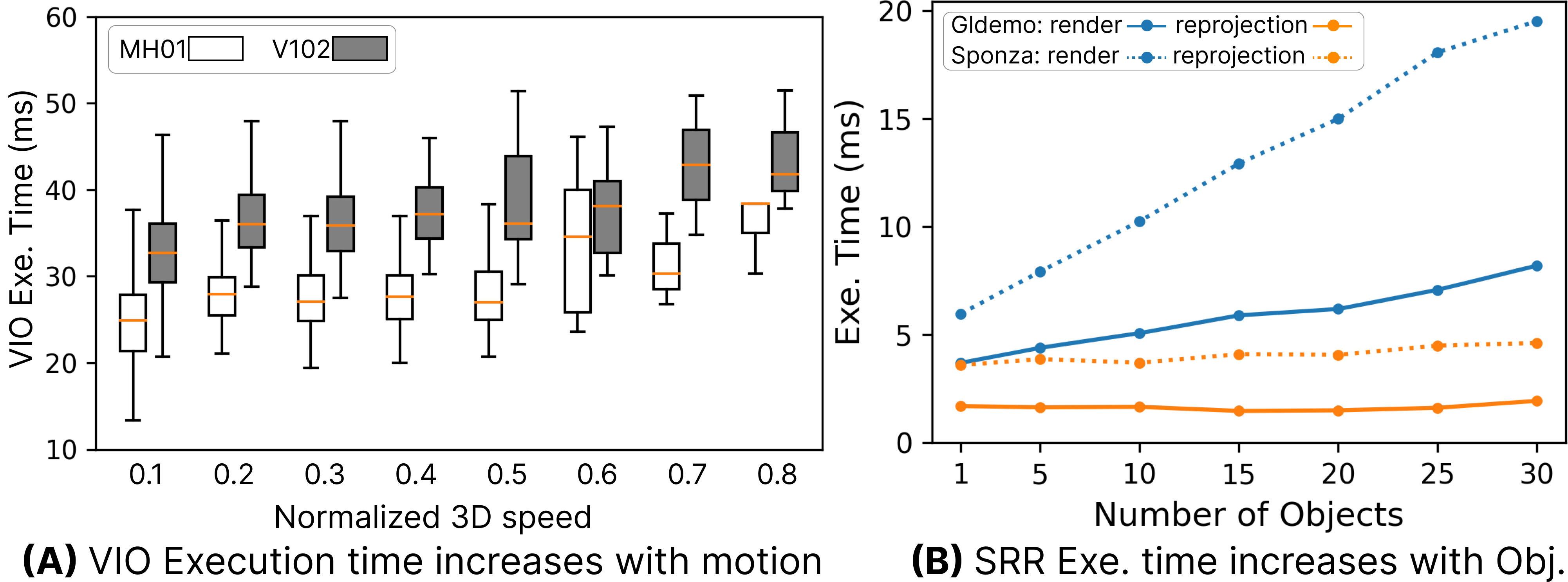}
\caption{XR characteristics under motion and scene dynamics.}
\label{fig_c1_c2}
\end{figure}

\textbf{C1: Motion-dependent Tracking.} During the perception phase, the XR system processes captured user motion through the VIO task, which exhibits significant variations in execution time when exposed to a broad range of motion. To analyze how motion dynamics affect VIO execution times, we test OpenVINS~\cite{openvins}, a widely used VIO algorithm that is standard in the ILLIXR framework, running on an embedded platform with two distinct trajectories from the EuRoc MAV dataset~\cite{euroc_mav}, specifically recorded in the Machine Hall (MH01) and Vicon Room (V102) environments. The execution time of VIO is recorded across a spectrum of normalized 3D speeds up to the maximum speed observed in the dataset. As shown in Fig.~\ref{fig_c1_c2}(A), VIO execution time increases with speed in both recorded trajectories. This escalation is due to the feature extraction front-end which detects more new features at higher speeds. Since VIO employs the MSCKF for the pose estimation part that calculates based on these new features, a larger number of features updated during extraction prolongs the execution time of the pose estimation, resulting in the overall increase of VIO execution time.

\textbf{C2: Scene-dependent Rendering.} During the visualization phase, the complete 3D frame needs SRR to render the 2D frame, which the ATWR reprojects. Therefore, we denote SRR execution time as render time and ATWR execution time as reprojection time. We investigate the correlation between the render and reprojection time when the objects in the viewport increase. We set up the experiment with only SRR, ATW, and ATWR running sequentially on the same two XR applications (Sponza and Gldemo) used in Sec.~\ref{subsec_optimization_space}. Throughout the experiment, we manipulate the number of objects by directly altering the objects in the viewport. As shown in Fig.~\ref{fig_c1_c2}(B), the render time increases linearly with the number of objects, whereas the reprojection time remains relatively constant. This discrepancy can be attributed to the workload heterogeneity of SRR and ATWR. SRR necessitates the rendering of every vertex, thereby requiring the calculation of each vertex's coordinates based on the given viewport. In contrast, ATWR reprojects only the updated vertices using the depth information from ATW without re-rendering the entire frame. Therefore, despite object changes in the scene, the impact on ATWR's execution time is minimal, whereas SRR's execution time increases linearly with the number of objects.

\begin{observation}
Even if the co-optimization of M2D and C2D is achieved, it can be easily nullified by motion dynamics (C1) and scene changes (C2). Therefore, XR systems need techniques to control the execution time fluctuations caused by these factors.
\label{obs_3}
\end{observation}

\section{Methodology}
\label{sec_methodology}

\subsection{BOXR Overview}

\begin{figure}[t]
\includegraphics[width=\linewidth]{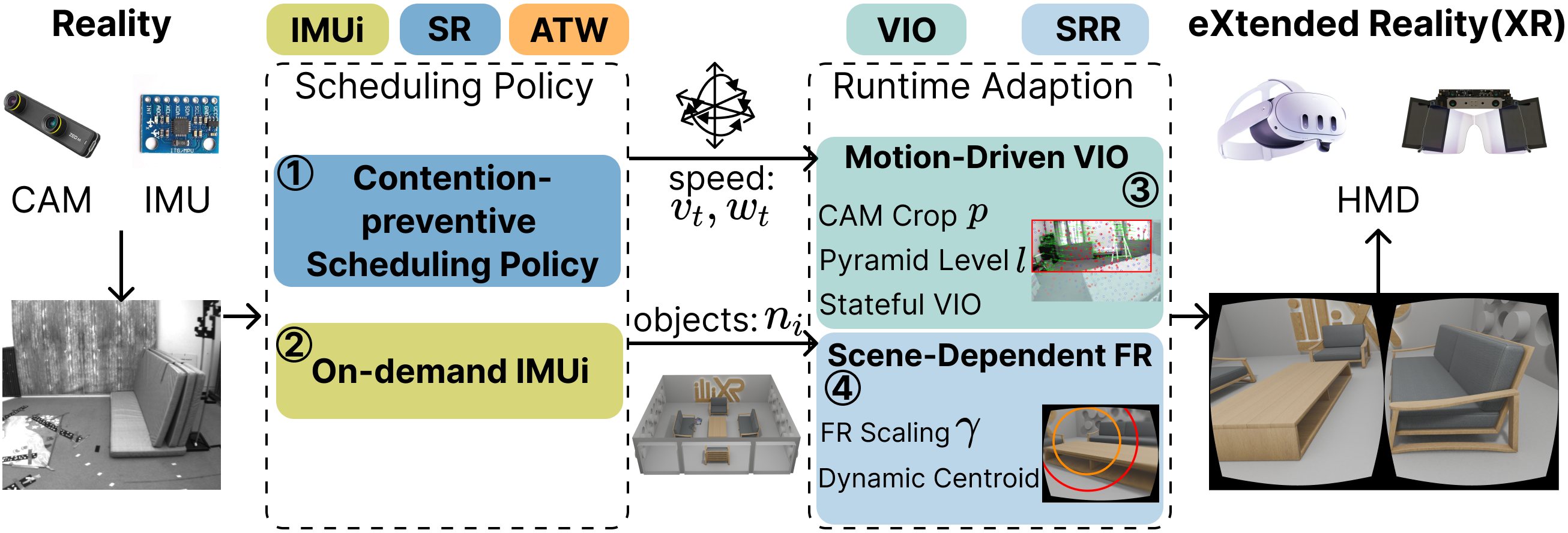}
\caption{BOXR Overview.}
\label{fig_overview}
\end{figure}

We present BOXR, a body and head motion optimization framework for XR based on the three observations from Sec.~\ref{sec_challenges} and show its system design in Fig.~\ref{fig_overview}.

Motivated by Obs.~\ref{obs_1} and Obs.~\ref{obs_2}, the contention and outdated raw pose result in large M2D and C2D, which can be addressed with a better scheduling policy design. BOXR proposes a contention-preventive scheduling policy (\circled{1} in Fig.~\ref{fig_overview}) that manages task execution sequences to maintain the up-to-date raw pose while preventing contentions. It further designs an on-demand IMUi (\circled{2} in Fig.~\ref{fig_overview}) to mitigate the problem of dropped IMU sensor information during the execution. Through the BOXR scheduling policy, we obtain an execution sweet spot given static execution time for all tasks.

Based on Obs.~\ref{obs_3}, the execution sweet spot derived from the BOXR scheduling policy can be nullified by the motion dynamics (C1) and scene changes (C2). Therefore, BOXR introduces the following runtime adaptions that each targets at an XR characteristic. To negate the effect of C1, BOXR designs a motion-driven visual inertial odometer (MVIO) (\circled{3} in Fig.~\ref{fig_overview}) that crops the input CAM image and controls the image pyramid level of the feature extraction based on the current motion magnitude. Also, to limit the extent of any resulting positional error, it includes an error bounding mechanism. BOXR then designs a scene-dependant foveated rendering (SFR) (\circled{4} in Fig.~\ref{fig_overview}) that replaces the original SRR to address C2. SFR dynamically changes the foveation area according to the number of objects in the viewport and centers the foveation area to the objects centroid.


\subsection{BOXR Scheduling Policy}
\label{subsec_boxr_scheduling_policy}

In this subsection, we assume the execution time of all tasks stays constant when designing the scheduling policy. This assumption will be lifted later in Sec.~\ref{subsec_mvio} and Sec.~\ref{subsec_sfr}. BOXR profiles the execution time for tasks VIO, IMUi, SR, SRR, ATW, and ATWR as $t_\text{VIO}$, $t_\text{IMUi}$, $t_\text{SR}$, $t_\text{SRR}$, $t_\text{ATW}$, and $t_\text{ATWR}$, respectively, and this needs to be done only once for a target hardware platform. To maintain the given sensor sample rate and target frame rate, BOXR assumes the period of CAM ($T_\text{CAM}$) which equals the period of VIO ($T_\text{VIO}$), the period of IMU ($T_\text{IMU}$), and the period of ATW ($T_\text{ATW}$) are given in advance. Without loss of generality, they follow $T_\text{CAM}>T_\text{ATW}\gg T_\text{IMU}$. The scheduling policy mentioned below is executed prior to each VIO execution.


\textbf{Contention-preventive Scheduling.} To address the challenges from Obs.~\ref{obs_1} and Obs.~\ref{obs_2}, we aim to eliminate the contention between SRR and ATWR while providing an up-to-date raw pose to ATWR.

At first, to prevent unnecessary contention between ATWR and SRR in any C2D sequence, we schedule the first SR job to start its execution immediately after the completion of the VIO job. 
This ensures the first SR job always receives the raw pose result from the latest VIO. In addition, within each $T_\text{VIO}$, the asynchronous execution of SRR and ATWR contributes to their contention. Therefore, we schedule ATW and ATWR jobs synchronously to SR and SRR jobs such that they begin only after the completion of the SRR job.

Next, to ensure the freshness of the raw pose data in the final output frame of ATWR, we co-optimize the SR \& SRR and ATW \& ATWR periods with the following approach. Since $T_\text{VIO}>T_\text{ATW}$, multiple output frames from ATWR exist in each $T_\text{VIO}$. If a different time interval from SR \& SRR is applied to ATW \& ATWR to process each frame, the SRR job from the previous C2D instance (VIO$\rightarrow$IMUi$\rightarrow$SR$\rightarrow$SRR$\rightarrow$ATW$\rightarrow$ATWR) may overlap with the ATWR job in the current C2D instance. This contributes to the outdated raw pose in Obs.~\ref{obs_2}, which inevitably results in large C2D variation. If an equal interval to execute each sequence is given, the ATWR and SRR overlapping can be safely avoided. Therefore, we first calculate the number of C2D instances within $T_\text{VIO}$ by $m=\lceil \frac{T_\text{VIO}}{T_\text{ATW}}\rceil$. We then equally divide $T_\text{VIO}$ by $m$ to determine the period of SR, $T_\text{SR}$, by $T_\text{SR}=\frac{T_\text{VIO}}{m}$. Although we do not explicitly change the period of ATW, $T_\text{ATW}$, it is implicitly determined since ATW runs synchronously to SR under our scheduling approach. To avoid additional contention caused by the delay of some ATW jobs, we further set a lower bound on $T_\text{SR}$, with $T_\text{SR} \geq t_\text{IMUi}+t_\text{SRR}+t_\text{ATW}+t_\text{ATWR}$. This eliminates the possibility of choosing a too small $T_\text{SR}$ since there will always be $t_\text{ATWR}$ time between two SRR jobs.

\begin{figure}[t]
\includegraphics[width=\linewidth]{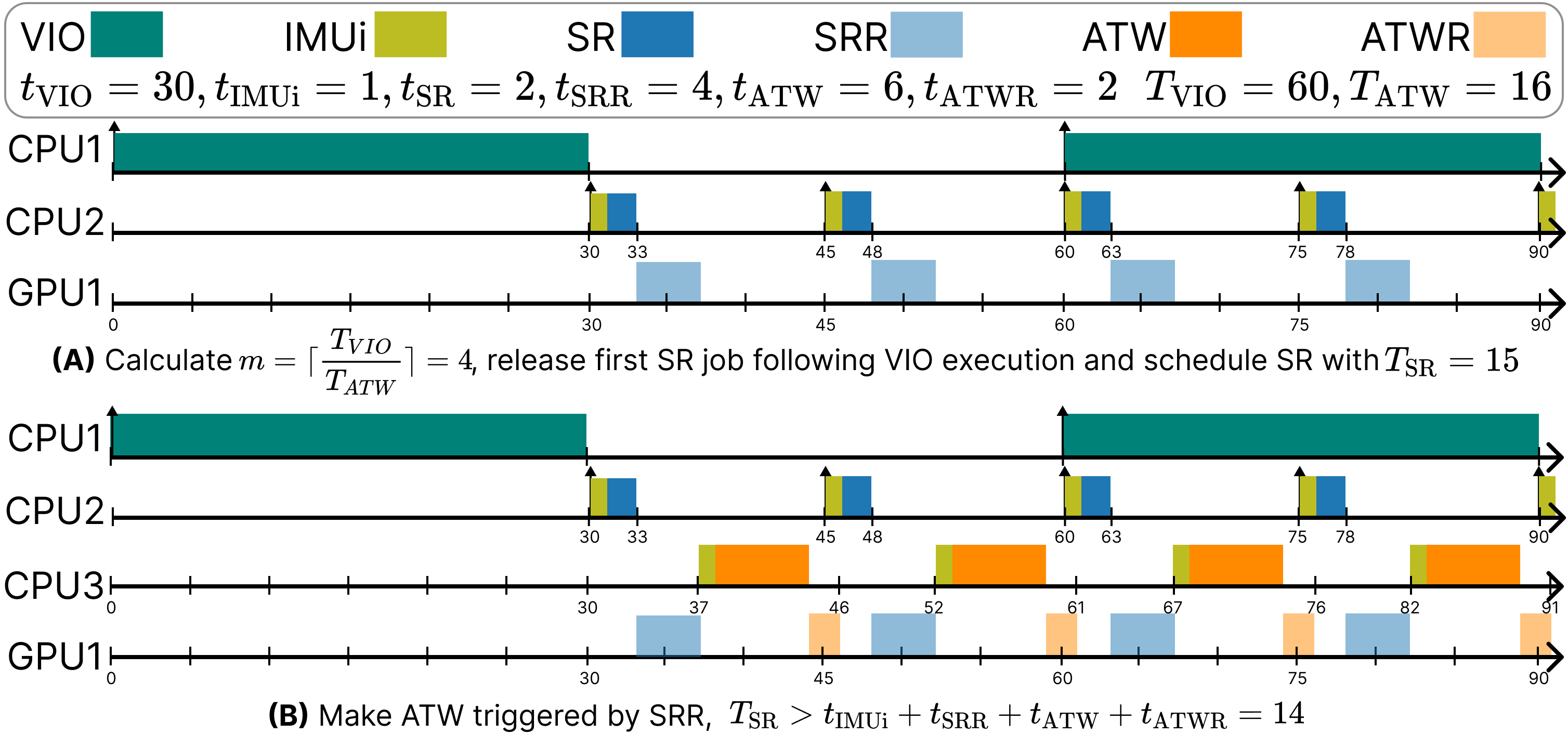}
\caption{BOXR scheduling example}
\label{fig_s_boxr_sched}
\end{figure}

\textbf{On-demand IMUi.} IMUi processes every IMU result to maximize the availability of the fused pose, which leads to wasted work of IMUi since the output fused pose is only used by SR and ATW, which satisfies $T_\text{SR} \simeq T_\text{ATW}\gg T_\text{IMU}$. Although decreasing the IMU sample rate proves effective in reducing wasted work, it delays the update of the fused pose, resulting in a significant increase in M2D and C2D.

To address this problem, we propose an on-demand IMUi method that samples the IMU reading and computes the fused pose at the beginning of each SR and ATW job. Since the fused pose is only needed before SR and ATW, this method keeps the same M2D and C2D with a timely fused pose update. 

We present the complete scheduling example in Fig.~\ref{fig_s_boxr_sched}. Fig.~\ref{fig_s_boxr_sched}(A) shows that given $T_\text{CAM}=T_\text{VIO}=60$ and $T_\text{ATW}=16$, we calculate $m=4$ in the first $T_\text{VIO}$. It then depicts the scheduling decisions for each VIO, IMUi, SR, and SRR job based on the contention-preventive scheduling and on-demand IMUi, where the first job of VIO is released at 0 and immediately followed by the sequential execution of IMUi, SR, and SRR. Contention-preventive scheduling further calculates the $T_\text{SR}=15$ and executes  IMUi, ATW, and ATWR upon the finish of SRR, as shown in Fig.~\ref{fig_s_boxr_sched}(B). The $T_\text{SR}$ is then compared with the designed lower bound to prevent contention.

\subsection{Motion-Driven Visual Inertial Odometer (MVIO)}
\label{subsec_mvio}

\begin{figure}[t]
\includegraphics[width=\linewidth]{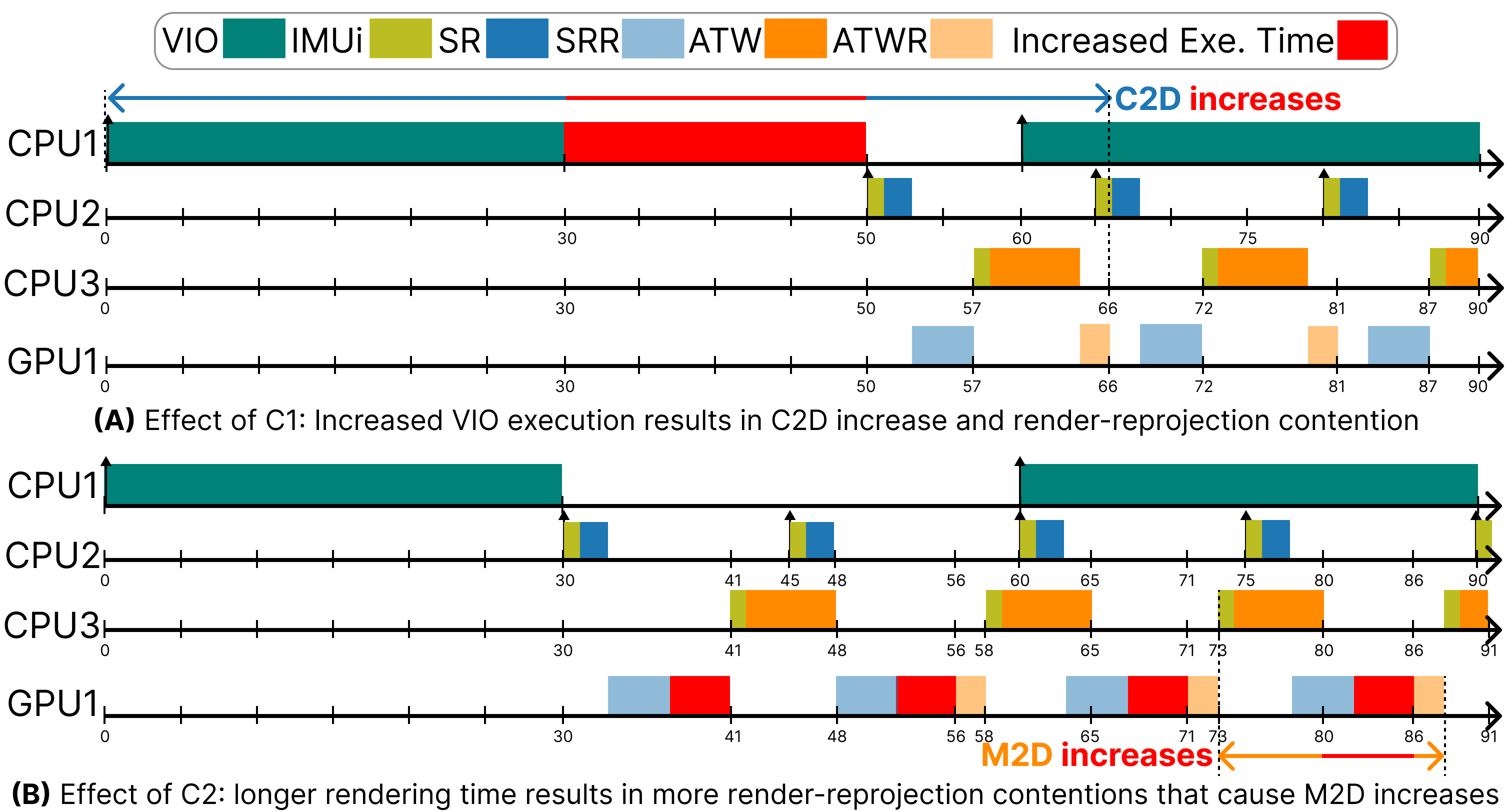}
\caption{Impact of C1 and C2 on scheduling}
\label{fig_d_boxr_c1_c2}
\end{figure}

\begin{figure}[t]
\includegraphics[width=\linewidth]{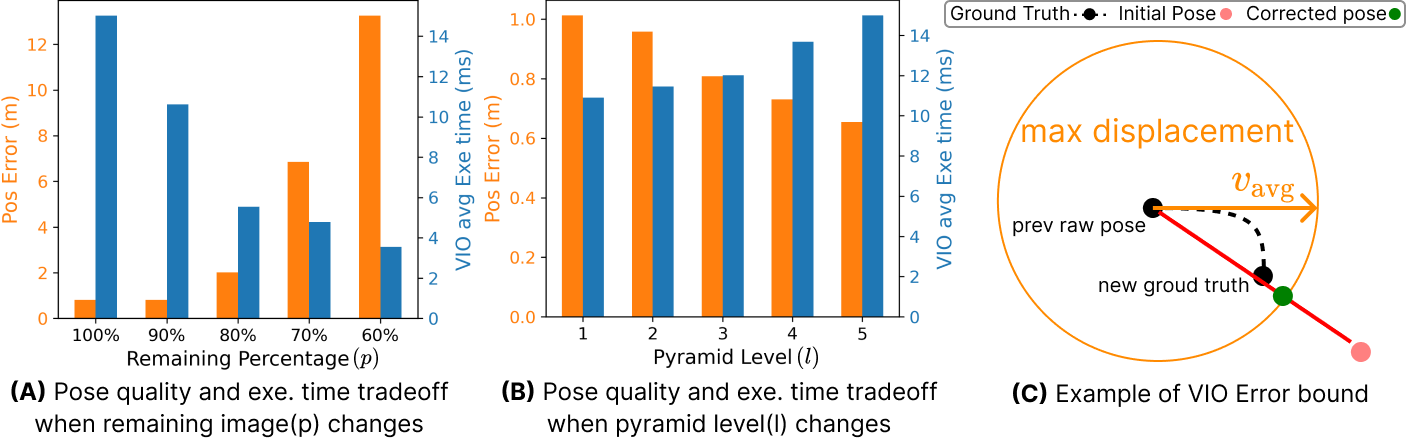}
\caption{Motion-Driven Visual Inertial Odometer}
\label{fig_c1_sol}
\end{figure}

As explained with C1 from Obs.~\ref{obs_3}, a larger magnitude of motion inflates the VIO execution time due to the increase of new features extracted. Fig.~\ref{fig_d_boxr_c1_c2}(A) illustrates its impact on the schedule. As the VIO execution time increases, the release of the first SR job for the current $T_\text{VIO}$ is delayed, leading to a significant increase in C2D. Therefore, we control the increase of VIO execution time by limiting the number of features through cropping the input image to VIO as well as decreasing the level of the pyramid for its feature pyramid network~\cite{openvins}. We denote the remaining input image percentage as $p$ and the pyramid level as $l$. To further analyze the effect of image crop and pyramid level adjustment, we run the VIO task on a PC with V102 trajectory used in Fig.~\ref{fig_c1_c2} and record the pose quality measured in position error and VIO execution time by only changing $p$ in Fig.~\ref{fig_c1_sol}(A) and by only changing $l$ in Fig.~\ref{fig_c1_sol}(B). We observe that as $p$ decreases, indicating a larger cropped area, the execution time decreases exponentially while the position error increases exponentially. As $l$ increases, indicating a higher pyramid level, the execution time increases linearly as the position error decreases linearly. 

Due to the tradeoff presented in Fig.~\ref{fig_c1_sol}, we design the Motion-Driven Visual Inertial Odometer (MVIO) to control the growth of execution time while limiting the decrease of pose quality. MVIO first sets a budget of VIO execution time $B_\text{VIO}=t_\text{VIO}$, where $t_\text{VIO}$ is the VIO execution time used in Sec~\ref{subsec_boxr_scheduling_policy}. It changes the remaining input image percentage $p$ and pyramid level $l$ and corrects any obvious position errors with its error bounding method. MVIO executes prior to each VIO job, with a complete algorithm provided in Alg.~\ref{alg_c1}.

\begin{algorithm}[t]
\footnotesize
\caption{MVIO Algorithm}\label{alg_c1}
{\textbf{Input:} IMU $a_t, v_t, w_t$, profiling data $v_{B}, w_{B}, f(e, l_\text{min})$, $e_\text{req}, S_\text{max}$\;}
/* Motion Score */\;
$t_{prev} \gets $ Timenow()\label{line_mvio_start}\;
$S\gets$ CalculateMotionScore($v_t, w_t, v_B, w_B$) /*Eq.~\ref{eq_motion_score}*/\label{line_motion_score}\;
$p\gets 1$\label{line_no_adj_start}\;
$l\gets l_\text{max}$\label{line_no_adj_end}\;
\If {$S>0$}{
/* Image Crop and Pyramid Level Adjustments */\label{line_adj_start}\;
$p_\text{min} \gets$ SetMinimumCrop($f(e_\text{req}, l_\text{min})$)\label{line_min_p}\;
$p \gets$ CalculateP($S, S_\text{max}, p_\text{min}$) /*Eq.~\ref{eq_p}*/ \label{line_calculate_p}\;
$l \gets$ ChangeL($S, S_\text{max}$) /*Eq.~\ref{eq_l}*/ \label{line_adj_end}\;
}
$image \gets $ CropImage(p, $a_t$) /*Crop image based on $a_t$ direction*/\label{line_crop_img}\; 
/* Start of VIO execution */\label{line_vio_start}\;
$features \gets$ FeatureExtraction($image, l$)\;
$pose \gets$ PoseEstimation($features$)\label{line_vio_end}\;
/* VIO Error Bounding */\label{line_stateful_vio_start}\;
$t_\text{VIO} \gets $ Timenow() - $t_{prev}$\;
$raw\_pose \gets$ readRawPose() /*Read latest $raw\_pose$*/\label{line_read_raw_pose}\;
$max\_dis \gets$ CalculateMaxDisplacement($v_{t}, a_{t}, t_\text{VIO}, raw\_pose$)\;
\If{$pose>max\_dis$}{
$pose \gets$ LinearFit($raw\_pose, max\_dis$)\label{line_stateful_vio_end}\;
}
$raw\_pose \gets pose$ /*Publish updated $raw\_pose$*/\;
\end{algorithm}

\textbf{Motion Score.} To capture the motion changes from the default speed and rotation, MVIO first computes a motion score $S$ which quantifies the deviation of the current motion to the representative motion from profiling. During the profiling of VIO execution time, we obtain the scalar value of velocity $v_B$ and scalar value of rotation $w_B$ when VIO completes execution in a given budget $B_\text{VIO}$ with the original image ($p=100\%$) and the max pyramid level $l_\text{max}$. 
At runtime, MVIO invokes IMU to get the acceleration and rotation at current time $t$, and calculates the first integral of $a_t$ to get the current scalar velocity $v_t$ and the current scalar rotation $w_t$. 
MVIO then constructs $S$ that describes the motion deviation from the $v_B$ and $w_B$, which is shown in Eq.~\ref{eq_motion_score}. During the profiling, we record the maximum observed velocity and rotation and use them to obtain the maximum possible motion score $S_\text{max}$, which is given as input to the MVIO algorithm.
\begin{equation}
     S=\frac{v_t-v_{B}}{v_{B}}+\frac{w_t-w_{B}}{w_B}
    \label{eq_motion_score}
\end{equation}

The MVIO algorithm given by Alg.~\ref{alg_c1} calculates $S$ at the beginning (line~\ref{line_motion_score}) as it is used in subsequent calculations. Initially, MVIO sets $p=1$ and $l=l_\text{max}$, which means no image cropping or pyramid level reduction from the maximum level (line~\ref{line_no_adj_start} and line~\ref{line_no_adj_end}). $S>0$ indicates a larger motion compared to the default $v_B$ and $w_B$ is detected, necessitating image cropping and pyramid level adjustment (line~\ref{line_adj_start} to line~\ref{line_adj_end}) to keep the execution time below $B_\text{VIO}$. Otherwise, MVIO maintains the default full image and the maximum pyramid level (line~\ref{line_no_adj_start} to line~\ref{line_no_adj_end}). 

\textbf{Image Crop Adjustment.} To limit the pose quality degradation, MVIO bounds the position error with the minimum remaining percentage $p_\text{min}$ that is calculated from a profiling function (line~\ref{line_min_p}). Let us denote the position error as $e$. We construct a function $p=f(e, l_\text{min})$ through profiling, which describes the relationship of $p$ to $e$ when applying the minimum pyramid level $l_\text{min}$, as shown in Fig.~\ref{fig_c1_sol}(B). By design, we allow the user to determine the maximum position error tolerable in the system, denoted by $e_\text{req}$. Therefore, by $f(e_\text{req}, l_\text{min})$, we get the minimum remaining percentage $p_\text{min}$, which is used during image cropping to limit the error to $e_\text{req}$. Note that, since the function $f(e, l_\text{min})$ has been profiled with $l_\text{min}$, the use of $p\ge p_{min}$ ensures at most $e_\text{req}$ positional error when adjusting $l$ later as long as $l\ge l_{min}$.

As shown by Fig.~\ref{fig_c1_sol}(B), VIO execution time and remaining image percentage $p$ have an exponential relationship. In addition, as illustrated in Fig.~\ref{fig_c1_c2}(A), VIO execution time has an approximately linear relationship to motion speed. Hence, we formulate the calculation of $p$ with an exponential relationship to $S$, given by Eq.~\ref{eq_p}.
\begin{equation}
    p = \max(p_\text{min}, p_{\text{min}}^{\frac{S}{S_{\text{max}}}})
    \label{eq_p}
\end{equation}
The max function in Eq.~\ref{eq_p} ensures if any abnormally large speed and rotation result in an $S>S_\text{max}$, MVIO always uses $p_\text{min}$ to control the position error below $e_\text{req}$. The term $\frac{S}{S_\text{max}}$ returns the percentage difference between current $S$ to $S_\text{max}$. Through the exponential calculation, MVIO returns the $p$ value (line~\ref{line_calculate_p}) and crops the image based on $p$ and direction of motion indicated by the vector value of acceleration $a_t$ from IMU (line~\ref{line_crop_img}). This occurs because new features will always emerge from the direction of motion as new image content appears from that direction.

\textbf{Pyramid Level Adjustment.} As shown in Fig.~\ref{fig_c1_sol}(A), VIO execution time and pyramid level follow a linear decrease relationship that is discrete since the pyramid level is an integer value. Because we profile $p=f(e, l_\text{min})$ with a minimum pyramid level $l_\text{min}$, the parameter $e$ after image crop should still satisfy $e\le e_\text{req}$, which gives the room for $l$ adjustment. Therefore, we decrease $l$ from $l_{max}$ with a step-wise function in Eq.~\ref{eq_l} and compute it after the calculation of $p$ (line~\ref{line_adj_end}). 
During the computation, $S_\text{max}$ is evenly distributed to $l_\text{max}-l_\text{min}+1$ segments. We calculate the current $S$ and then compare it with each segment to determine the corresponding $l$ using the step-wise function. Through the process, MVIO can equally leverage the image crop and pyramid level reduction to control the growth of VIO execution time within the designed $B_\text{VIO}$ while maintaining the minimum position error $e_\text{req}$.

\begin{equation}
l = \begin{cases} 
l_{\text{max}}, & \text{if } S \leq S_{\text{max}} / l_{\text{max}} \\
l_{\text{max}}-1, & \text{if } S_{\text{max}} / l_{\text{max}} < S \leq (l_\text{min}+1)S_{\text{max}} / l_{\text{max}} \\
\hdots & \\
l_{\text{min}}, & \text{if } S > S_{\text{max}} / l_{\text{min}}
\end{cases}
\label{eq_l}
\end{equation}
 
 \textbf{VIO Error Bounding.}
 Due to the introduction of $e_\text{req}$, MVIO further employs a VIO error bounding method to limit the position error by leveraging the raw pose published by the previous VIO. After the execution of the VIO algorithm (line~\ref{line_vio_start} to line~\ref{line_vio_end}), MVIO calculates the current VIO execution time $t_\text{VIO}$ that can be used to determine the maximum possible displacement from the previous raw pose result (line~\ref{line_read_raw_pose}). MVIO chooses to use the raw pose because the fused pose introduces further noise from the head motion during the IMUi calculation. The maximum displacement $max\_dis$, indicated as the orange circle in Fig.~\ref{fig_c1_sol}(C), is computed by getting the average velocity $v_\text{avg}$ during $t_\text{VIO}$ with $v_\text{avg}=v_t+a_t\times t_\text{VIO}$ and then apply this speed to the previous raw pose by $max\_dis = raw\_pose+v_\text{avg}\times t_\text{VIO}$. If the $pose$ from the MVIO execution exceeds the $max\_dis$, the $pose$ is fitted to the $raw\_pose$ by finding the intersection between the line that connects $pose$ and $raw\_pose$ and the $max\_dis$.

\subsection{Scene-Dependent Foveated Rendering (SFR)}
\label{subsec_sfr}

\begin{figure}[t]
\includegraphics[width=\linewidth]{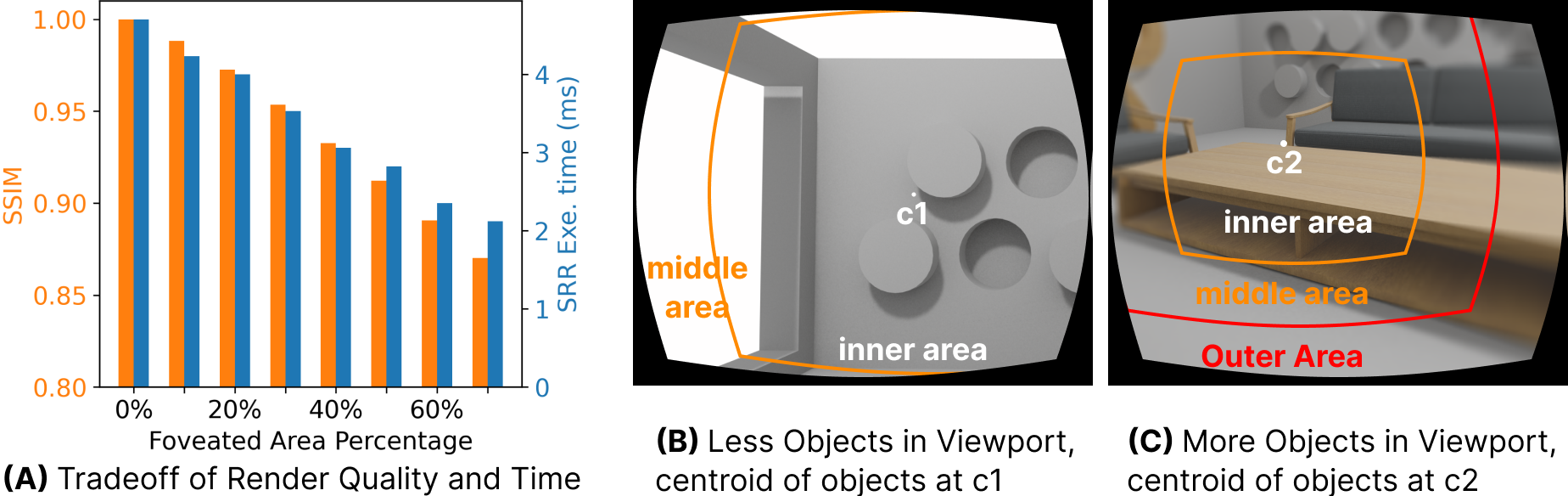}
\caption{Scene-Dependent Foveated Rendering}
\label{fig_c2_sol}
\end{figure}

Based on C2 from Obs.~\ref{obs_3}, we observe the increase of objects in the viewport causes the execution time of SRR (render time) to grow linearly, leading to the contention between SRR and ATWR again shown in Fig.~\ref{fig_d_boxr_c1_c2}(B). Existing work has developed the foveated rendering method that can reduce the rendering time by using low-resolution texture and meshes in the peripheral area~\cite{weier2016foveated,fujita2014foveated,weier2018foveated,kim2019foveated,lee2020foveated,jabbireddy2022foveated}. Foveated rendering typically involves three levels of rendering resolution: an inner area with the original resolution, a middle area with half the original resolution, and an outer area with one-quarter of the original resolution. We denote the inner area as the foveation area. While existing work uses a fixed foveation area that is not adaptive to scene changes, we change the foveation area and plot the tradeoff of frame quality measured in the Structural Similarity (SSIM) Index and render time in Fig.~\ref{fig_c2_sol}(A) with a fixed objects number. 
We choose to use the SSIM index because it measures the human-perceived difference in image, which is widely used in existing research~\cite {ssim_paper,illixr,ke2023collabvr}. Fig.~\ref{fig_c2_sol}(A) indicates both the quality of rendering and rendering time follow a linear decrease relationship to the area of foveation.

Motivated by this tradeoff, we design the Scene-Dependent Foveated Rendering (SFR) that aims to control the increase of rendering time with minimum degradation of rendering quality. SFR first adopts a budget of render time $B_\text{SRR}=t_\text{SRR}$. We then profile the number of objects that result in $B_\text{SRR}$ render time with no usage of foveated rendering and denote it as $n_{B}$. SFR replaces the default SRR and follows Alg.~\ref{alg_c2}.

\begin{algorithm}[t]
\footnotesize
\caption{SFR Algorithm}\label{alg_c2}
{\textbf{Input:} SR $vp\_objects$, profiling data $K, M, C, r_{max}, n_{B}$\;}
$n\gets$Lengthof($vp\_objects$) /*get number of objects in viewport*/\label{line_get_n}\;
$\alpha \gets 1$\label{line_alpha_1}\;
/* Foveation Scaling Factor */\;
\If{$n>n_{B}$}{
$\gamma \gets$ CalculateGamma($\alpha, K,M,C,n_{B}$)\label{line_calculate_gamma} /*Eq.~\ref{eq_sfr_gamma}*/\;
/* Optimization Loop */\label{line_opt_loop_start}\;
    \While{$C\gamma>B_\text{SRR}$}{ 
    \If{$\alpha > 0$}{
        $\alpha \gets \alpha-0.1$\;
        $\gamma \gets$ CalculateGamma($\alpha, K,M,C,n_{B}$) /*Eq.~\ref{eq_sfr_gamma}*/\label{line_opt_loop_end}\;
    }
    }
}
/* Foveated Rendering */\;
$inner\_area\gets \gamma \cdot (height \times width$)\label{line_sft_set_start}\;
$middle\_area\gets(\gamma\cdot height+offset)\times(\gamma\cdot width+offset)-inner\_area$\;
$outer\_area\gets outer\_area-middle\_area-inner\_area$\label{line_sft_set_end}\;
$c\gets$CalculateCentroid($vp\_objects$) /*get centroid of objects*/\label{line_calculate_centriod}\;
$2D\_frame\gets$ FoveatedRender($c, inner\_area, middle\_area, outer\_area$)\;
$frame\gets 2D\_frame$ /*publish rendered 2D frame*/\;
\end{algorithm}

\textbf{Foveation Scaling Factor.} Due to the linearity of render quality and render time to the foveation area in Fig.~\ref{fig_c2_sol}(A), we set up a foveation scaling factor, denoted as $\gamma$, to scale the area of foveation which limits the growth of render time within the designed $B_\text{SRR}$. Through offline profiling, we are able to get the linear coefficients of each linear relationship. In the following equations, $C$ denotes the linear coefficient of render time to $\gamma$, $M$ denotes the linear coefficient of rendering quality to $\gamma$, and $K$ denotes the linear coefficient of render time to number of objects in the viewport. To start up, the objects in the current frame's viewport, denoted as $vp\_objects$, are provided alongside $C$, $K$, and $M$. SFR first gets the number of $vp\_objects$ (line~\ref{line_get_n}) $n$ and calculates the $\gamma$ according to Eq.~\ref{eq_sfr_gamma} when $n>n_{B}$ (line~\ref{line_calculate_gamma}). In the equation, $\alpha$ is used as a control value that is initially set to 1 to max out the frame quality (line~\ref{line_alpha_1}). The first term indicates that when $\gamma=1$, no foveated rendering is used. The next term $\frac{(1 - \alpha)M}{\alpha C}$ indicates the contributing factor from render quality. The last term $\frac{(n - n_B)}{\alpha KC}$ indicates the contributing factor from object number changes.

\begin{equation}
    \gamma = 1 - \frac{(1 - \alpha)M}{\alpha C} - \frac{(n - n_B)}{\alpha KC}
    \label{eq_sfr_gamma}
\end{equation}

\textbf{Optimization Loop.} With the initial $\gamma$ changes, it is still possible that the rendering time exceeds the $B_\text{SRR}$, so we opt to further reduce the rendering time by making concessions in rendering quality with a decreased $\alpha$ in Eq.~\ref{eq_sfr_gamma}. This process is called the Optimization loop and is presented in line~\ref{line_opt_loop_start} to line~\ref{line_opt_loop_end}. Initially, SFR projects the rendering time with the calculation of $C\gamma$ and compares it with the $B_\text{SRR}$. If the projected rendering time exceeds $B_\text{SRR}$, SFR lowers the value of $\alpha$ by $0.1$ and recalculates $\gamma$ using Eq.~\ref{eq_sfr_gamma}. This loop ends when $\alpha=0$ or $C\gamma \leq B_\text{SRR}$. 

\textbf{Dynamic Objects Centroid.} Since the centroid of objects does not align with the viewport center, we need to calculate the objects' centroid and fix the center of foveation there to effectively reduce the rendering time. Therefore, after setting the inner area, middle area, and outer area (line~\ref{line_sft_set_start} to line~\ref{line_sft_set_end}), SFR calculates the centroid of objects in the viewport by taking the average of all objects geometry center x coordinates $x_\text{avg}$ and y coordinates $y_\text{avg}$ and returns the centroid coordinates as $(x_\text{avg}, y_\text{avg})$. We show two examples of 2D frames rendered by SFR in Fig.~\ref{fig_c2_sol}(B). 

\subsection{Generalizability}
Since BOXR only modifies the scheduler and three tasks (VIO, IMUi, and SRR) of the state-of-the-art XR framework, it can be directly applied to any XR systems that share similar tasks and pipelines. When deploying on a new platform, BOXR requires re-profiling task execution times for a target hardware setup. This step provides the execution times needed for the BOXR scheduling policy and sets the budget for MVIO and SFR. To eliminate the manual effort involved in the process, BOXR includes a setup application that collects all necessary execution times for each task. These data are then fed into our scheduler, MVIO, and SFR algorithm to determine the initial conditions for execution. Following the same process, we have successfully deployed BOXR onto three different hardware platforms in Sec.~\ref{subsec_evaluation_setup}.

One may have a concern that since the execution times of all tasks have to be re-profiled for each hardware setup, it may generate profiling errors that can interfere with the scheduling decisions and lead to performance degradation. 
However, since the MVIO and SFR use the same profiling data to make runtime adaption, they can offset the inaccuracy by reactively changing the magnitude of adjustment. For example, if a shorter VIO execution time is profiled, MVIO will crop the image more aggressively due to a larger difference between $B_\text{VIO}$ and $t_\text{VIO}$. Hence, VIO is still controlled below $B_\text{VIO}$ amidst motion dynamics.

\section{Evaluation}
\label{sec:eval}
\subsection{Evaluation Setup}
\label{subsec_evaluation_setup}

\textbf{Hardware Setup}. To encompass a broad range of hardware platforms commonly used to support XR systems, we evaluate BOXR on three hardware platforms, including a PC equipped with NVIDIA GTX 3060 GPU and two embedded devices, NVIDIA AGX Xavier~\cite{xavier} and NVIDIA Orion Nano~\cite{nano}, which are widely used in cutting-edge XR research~\cite{jiang2023offloading,singh2023power,you2022eyecod,illixr}. We fix the power of the PC to 180W, Xavier to 30W, and Nano to 15W to control the GPU frequency and lock the CPU frequency of each embedded platform to its maximum value. The output frame is displayed on Northstar Next HMD~\cite{northstar_next} with a max 90Hz refresh rate.

\begin{table}[t]
\begin{center}
\caption{Evaluated XR Applications}
\scriptsize
\begin{tabular}{|c|c|c|c|c|}
\hline
& \textbf{Sponza}(Spon) & \textbf{Materials}(Mat) & \textbf{Gldemo}(Gl) & \textbf{Platformer}(Plat)\\
\hline
Object & 32 & 81 & 7 & 3014 \\
\hline
Vertex & 192870 & 62826 & 54760 & 26168 \\
\hline
Texture & 33 & 24 & 8 & 4 \\
\hline
\end{tabular}
\label{tab_xr_app}
\end{center}
\end{table}

\begin{table}[t]
\begin{center}
\caption{EuRoc MAV Dataset Categorization}
\scriptsize
\begin{tabular}{|c|c|c|c|c|}
\hline
& \textbf{No Motion} & \textbf{Small Motion} & \textbf{Med. Motion} & \textbf{Large Motion} \\
\hline
$v_{3D}(m/s)$ & $[0, 0.1)$ & $[0.1, 1)$ & $[1, 2)$ & $[2, \infty)$ \\
\hline
Perc$(\%)$ & 10.62 & 66.75 & 20.30 & 2.33 \\
\hline
\end{tabular}
\label{tab_dataset}
\end{center}
\end{table}

\textbf{XR Applications Setup.} To encompass diverse scene complexities that contribute to different rendering pressures in the everyday use case of XR, we selected four OpenXR~\cite{openxr} applications initially built with the Godot engine~\cite{godot}. Their details are shown in Table~\ref{tab_xr_app} and ordered by decreasing rendering demands. Sponza brings the user to a detailed palace for exploration. Materials renders a series of identical objects with various textures. Gldemo simulates an indoor scene resembling a household living room. Platformer creates a pixel-style world with numerous individual objects for gaming scenarios. We fixed the output resolution to $2560 \times 1440$ and set both horizontal and vertical fields of view to $90^{\circ}$. To avoid additional overhead from application interfaces and focus on the effects of motion and scene dynamics, we loaded only the scene information, including meshes and textures, from these applications and used OpenGL as the rendering engine for all rendering and reprojection tasks.

\textbf{Controlled Trajectories.} We controlled the input visual-inertial information using the eleven trajectories from EuRoc MAV datasets~\cite{euroc_mav}. This dataset covers the majority of indoor XR use cases with average speed from $0.33m/s$ to $0.99m/s$ and rotation from $0.21rad/s$ to $0.66rad/s$. We set the CAM period to 50ms and the IMU period to 5ms which is used by the dataset. To understand the effect of head and body motion on the system performance, we categorized all the visual-inertial information into four motion classes shown in Table~\ref{tab_dataset} based on their ground-truth 3D speed ($v_{3D}$), which is the scalar value of the linear speed in 3D space composed of both velocity and rotation.

\begin{figure}[t]
\includegraphics[width=\linewidth]{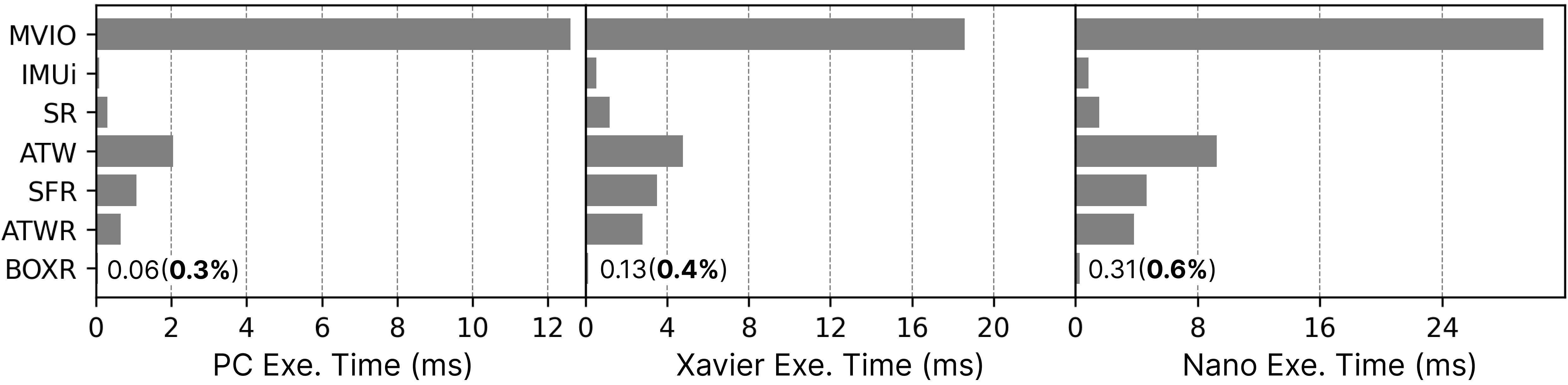}
\caption{System runtime breakdown}
\label{fig_overhead_breakdown}
\end{figure}

\textbf{Baseline Configuration}.
We compared BOXR against two state-of-the-art baselines derived from the ILLIXR framework~\cite{jiang2023offloading,singh2023power,liu2023demystifying,you2022eyecod}.  ILLIXR~\cite{illixr} is a popular open-sourced XR software testbed that is being widely used in XR research. 
\textit{ILLIXR} uses the default publisher-subscriber model in Sec.~\ref{subsec_xr_system_model} to schedule tasks, with V-Sync disabled to maximize throughput. \textit{ILLIXR-OP} uses the same optimized version of ILLIXR in Sec.~\ref{subsec_optimization_space}, which modifies the periods of SR and ATW to minimize M2D and C2D while still adhering to the same publisher-subscriber model. To assess the contribution of each component, we evaluate the scheduling policy and the complete framework separately by making them into two separate baselines. \textit{BOXR-S} is a static version that only implements the scheduling policy described in Sec.~\ref{subsec_boxr_scheduling_policy}. \textit{BOXR} implements the complete framework that includes the scheduling policy, MVIO, and SFR in Sec.~\ref{sec_methodology}. We fixed the target frame rate to 90 FPS, which matches the max screen refresh rate, and recorded the results of M2D and C2D at the end of each frame.



\begin{figure*}[t]
\includegraphics[width=\textwidth]{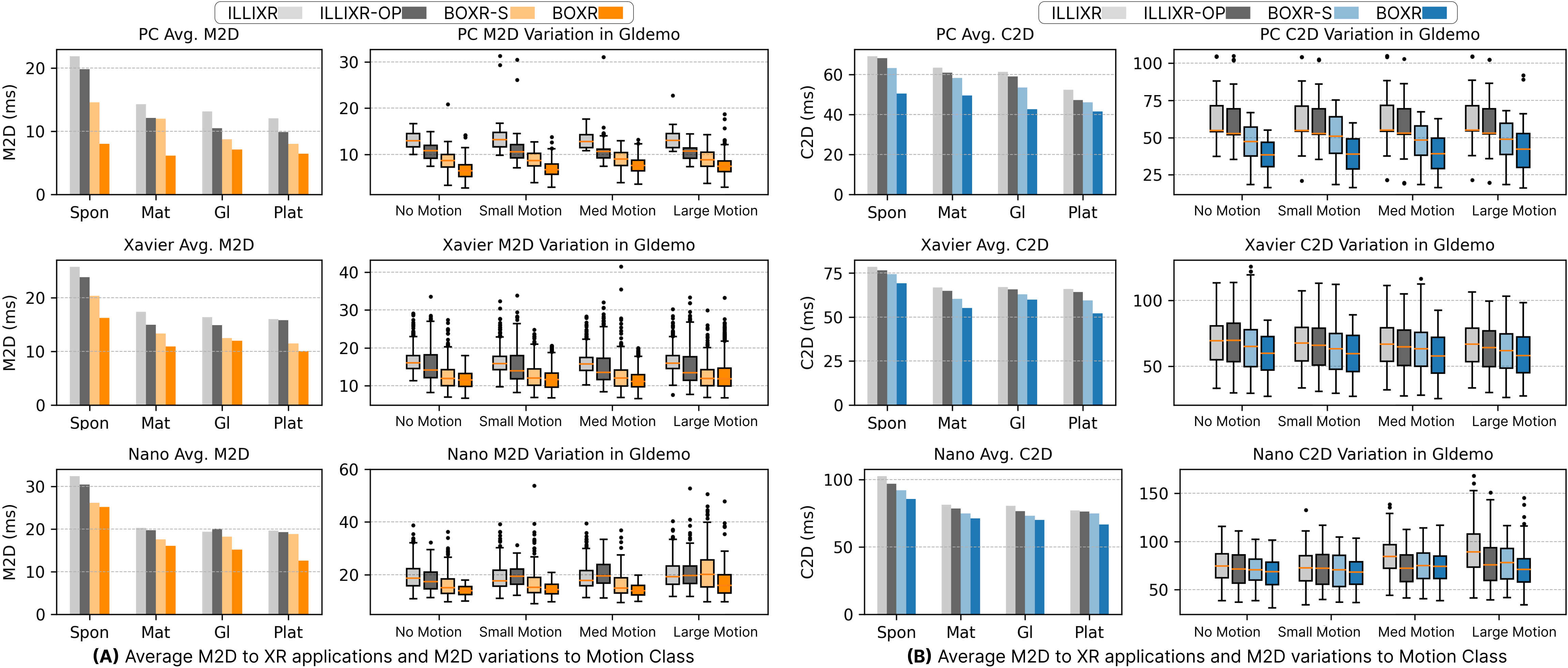}
\caption{M2D and C2D Evaluation. The plots from left to right show the average M2D, M2D variations, average C2D, and C2D variations. The M2D and C2D variations show each M2D and C2D given different motion classes in the Gldemo application.}
\label{fig_eval_m2d_c2d}
\end{figure*}

\subsection{Controlled Settings Effectiveness}


\textbf{Latency and Overhead Breakdown}. To evaluate the overhead of BOXR, we profiled the average execution time of each task in the XR framework. We use the MVIO to replace the VIO task and SFR to replace the SRR task. Within these tasks, we separately profile the overhead of running BOXR algorithms, including the scheduling policy, line~\ref{line_mvio_start} to line~\ref{line_crop_img} and line~\ref{line_stateful_vio_start} to line~\ref{line_stateful_vio_end} in Alg.~\ref{alg_c1}, as well as line~\ref{line_get_n} to line~\ref{line_calculate_centriod} in Alg.~\ref{alg_c2}. As shown in Fig.~\ref{fig_overhead_breakdown}, the average overhead of our method is around 0.06ms, 0.13ms, and 0.31ms compared to the entire average system runtime, which is more than 17ms on PC, 36ms on Xavier, and 48ms on Nano. The overhead of using our method is about 0.3\%, 0.4\%, and 0.6\%, respectively.

\begin{table}[t]
\begin{center}
\caption{Dropped IMU sensor information}
\scriptsize
\begin{tabular}{|c|c|c|c|c|}
\hline
& \textbf{MH01} & \textbf{MH05} & \textbf{V101} & \textbf{V102}\\
\hline
ILLIXR & 36\% & 26\% & 27\% & 35\% \\ \hline
ILLIXR-OP & 62\% & 46\% & 26\% & 73\% \\ \hline
BOXR-S & 0\% & 0\% & 0\% & 0\% \\ \hline
BOXR & 0\% & 0\% & 0\% & 0\% \\ \hline
\end{tabular}
\label{tab_dropped_imu}
\end{center}
\end{table}

\textbf{Dropped IMU Sensor Information.} We record the dropped IMU sensor information in two trajectories from the Machine Hall (MH01, MH05) and two from the Vicon Room (V101, V102) in EuRoC MAV. As shown in Table~\ref{tab_dropped_imu}, in ILLIXR, the IMU operates at a significantly shorter period than SR and ATW, leading to up to 36\% dropped information. ILLIXR-OP further increases the SR period, causing up to 73\% IMU sensor information to be dropped. BOXR-S, benefiting from On-demand IMUi, triggers IMU sensor readings only at the beginning of SR and ATW, resulting in no dropping of sensor information. Additionally, BOXR samples IMU readings at the beginning of VIO, which are used to calculate the raw pose leading to no drop of sensor information.

\begin{figure}[t]
\includegraphics[width=\linewidth]{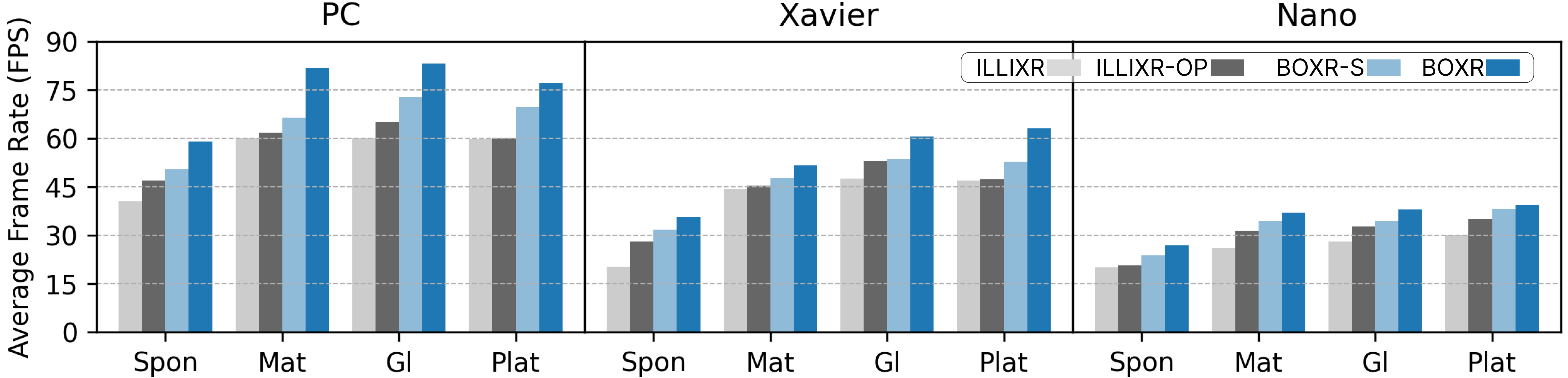}
\caption{Avg. output frame rate with 90FPS target frame rate}
\label{fig_eval_fps}
\end{figure}

\textbf{M2D Metric.} To obtain the M2D value defined in Def.~\ref{def_m2d}, we record the age of the latest IMU information at the time of frame output. The resulting average M2Ds are shown in the left subfigures of Fig.~\ref{fig_eval_m2d_c2d}(A). Compared to ILLIXR, BOXR-S reduces M2D by up to $33\%$ because the proposed scheduling policy effectively reduces contention between SRR and ATWR. Furthermore, BOXR achieves M2D reductions by up to $63\%$ on PC, $37\%$ on Xavier, and $36\%$ on Nano, thanks to SFR which further controls the rendering time when rendering a large number of objects.

The right subfigures of Fig. \ref{fig_eval_m2d_c2d}(A) demonstrate the variation of M2D. BOXR-S increases M2D standard deviation by $15\%$ compared to ILLIXR on Nano. As shown in Fig.~\ref{fig_d_boxr_c1_c2}(B), BOXR-S assumes constant task execution time, which cannot eliminate contention due to the dynamics brought by objects in the scene (C2) and potentially increases the M2D variation during runtime. However, BOXR reduces M2D standard deviation by up to $57\%$ compared to ILLIXR. This is due to the use of SFR, which bounds rendering time to avoid contention while maintaining similar quality by dynamically adjusting the foveation area and object centroid based on the number of objects in the scene.


\textbf{C2D Metric.} We add the CAM timestamp to each SRR-generated 2D frame and then record the age of this CAM information during the final 3D frame output as C2D. The left subfigures of Fig.~\ref{fig_eval_m2d_c2d}(B) illustrate the average C2D. BOXR-S achieves up to $13\%$ C2D reduction compared to ILLIXR since the scheduling policy effectively prevents contentions between SRR and ATWR. BOXR reduces C2D by up to $27\%$ compared to ILLIXR, owing to MVIO which bounds the VIO execution time across from motion dynamics.

The right subfigures of Fig.~\ref{fig_eval_m2d_c2d}(B) show the variation of C2D. BOXR-S experiences $6\%$ increase of C2D standard deviation compared to ILLIXR on Xavier. As explained in Fig.~\ref{fig_d_boxr_c1_c2}(A), the dynamics of motion (C1) contribute to variation increase since BOXR-S neglects the motion effect on VIO execution time. However, BOXR achieves up to $23\%$ less C2D standard deviation compared to ILLIXR. This significant improvement is made possible by MVIO, which controls the VIO execution time with image crop and pyramid level adjustment during large motion. Notably, the only extremely large C2D outliers for BOXR occur in the large motion class, which constitutes just $2\%$ of the entire dataset. This is because when unexpectedly large motion dynamics happen (i.e. sudden drop or turn), MVIO may still use the previous velocity to calculate the first VIO job during such abrupt motion dynamics, creating a single large C2D outlier. We will provide more evaluation about this in Sec.~\ref{subsec_burst_dynamics_scenarios}.


\begin{figure}[t]
\includegraphics[width=\linewidth]{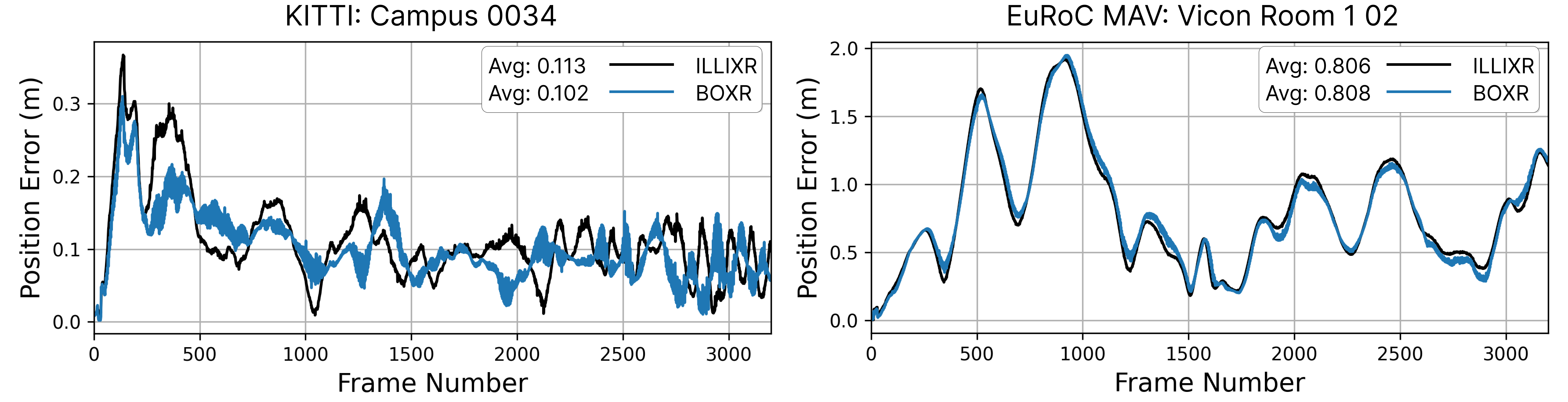}
\caption{Position error and render time-quality tradeoff}
\label{fig_eval_pos_err}
\end{figure}

\textbf{Average Output Frame Rate}. Figure~\ref{fig_eval_fps} shows the average output frame rate measured in frames per second (FPS) during a 60-second runtime, with the target frame rate fixed at 90 FPS. Due to reduced contention between SRR and ATWR through BOXR scheduling policy, BOXR-S achieves up to a 27\% frame rate increase compared to ILLIXR and up to a 16\% increase compared to ILLIXR-OP. BOXR achieves up to a 43\% frame rate increase compared to ILLIXR and up to a 35\% increase compared to ILLIXR-OP, thanks to the use of SFR that adapts to the objects change. With BOXR optimization, the PC achieves a consistent 60 FPS across all four applications, ensuring an ideal XR experience. Xavier achieves 60 FPS in Gldemo and Platformer, making it practical for low-render demand XR applications. Furthermore, BOXR enables usage of the XR system on Nano, which outputs 30 FPS for three applications.


\textbf{Quality of Pose.} To test the general effectiveness of the MVIO algorithm, we run both ILLIXR and BOXR with the Gldemo application on PC and record the position error from ground truth in Fig.~\ref{fig_eval_pos_err} for a one-minute trial. To differentiate the environments for VIO and evaluate the general applicability, we conduct the experiment with the KITTI~\cite{Geiger2013IJRR} camera feed for outdoor large motion scenario as well as the EuRoC MAV~\cite{euroc_mav} camera feed for indoor moderate motion scenario. BOXR decreases the position error by 9.7\% for the outdoor scenario of KITTI but yields a similar error for the indoor scenario of EuRoC MAV. 
Since KITTI involves significantly larger velocities, ILLIXR shows a larger variability of position errors due to the prolonged VIO execution time, resulting in unfinished VIO jobs. Benefiting from MVIO, BOXR mitigates the problem by controlling VIO execution time to avoid VIO jobs being dropped and achieve more accurate localization. Another observation is that BOXR accumulates position error more rapidly than ILLIXR when consecutively large motion dynamics happen, as shown in 1000 to 1500 frames. However, it quickly corrects the subsequent poses due to the use of VIO error bounding and reaches the same pose quality when motion magnitude decreases. This experiment shows that BOXR can maintain similar pose quality during indoor moderate motion scenarios, while even improving the pose quality in outdoor large motion scenarios.

\textbf{Rendering Output Quality.} Table~\ref{tab_quality_of_rendering} records the change of render time and quality of output frame measured in the Structural Similarity (SSIM) Index from ILLIXR to BOXR. SSIM captures the perceived quality of the output frame, which is a direct implication of user notice of visual change~\cite{ssim_paper}. Although we see a decrease of SSIM up to 5.7\% in Nano, we trade off this rendering quality degradation to a 30.4\% decrease in rendering time, which contributes to better M2D and C2D while increasing output frame rate.

\begin{table}[t]
\centering
\caption{Quality of Rendering for Different Platforms}
\scriptsize
\begin{tabular}{|l|c|c|c|c|c|c|}
\hline
& \multicolumn{2}{c|}{\textbf{PC}} & \multicolumn{2}{c|}{\textbf{Xavier}} & \multicolumn{2}{c|}{\textbf{Nano}} \\
\cline{2-7}
& $\Delta$Time & $\Delta$SSIM & $\Delta$Time & $\Delta$SSIM & $\Delta$Time & $\Delta$SSIM \\
\hline
Spon & -15.0\% & -3.7\% & -25.4\% & -5.6\% & -30.4\% & -5.7\% \\
Mat & -16.0\% & -2.7\% & -21.6\% & -4.6\% & -26.1\% & -4.8\% \\
GL & -6.5\%  & -1.2\% & -16.9\% & -1.8\% & -22.3\% & -3.4\% \\
Plat & -12.9\% & -1.9\% & -10.5\% & -1.3\% & -20.8\% & -3.0\% \\
\hline
\end{tabular}
\label{tab_quality_of_rendering}
\end{table}

\begin{minipage}{0.92\linewidth}
\begin{shaded}
    \noindent \textbf{Overall Effectiveness:} BOXR enhances performance across all platforms by addressing the challenges presented in Obs.~\ref{obs_1} and Obs.~\ref{obs_2} through its scheduling policy design and tackling the challenge from Obs.~\ref{obs_3} with runtime adaptation.
\end{shaded}
\end{minipage}

\subsection{Burst-dynamics Scenarios Performance}\label{subsec_burst_dynamics_scenarios}

\begin{figure}[t]
\includegraphics[width=\linewidth]{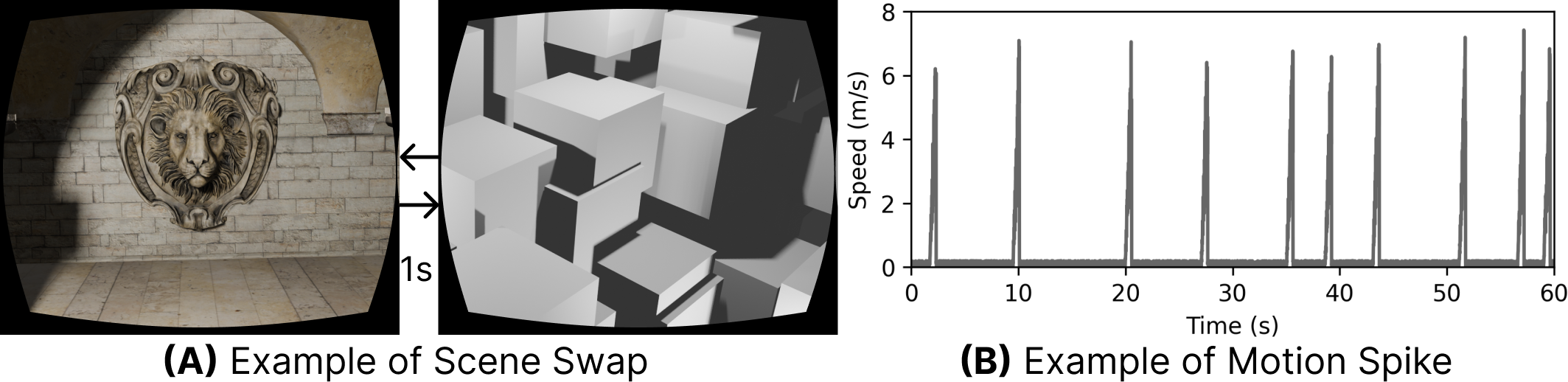}
\caption{Scene swap and motion spike}
\label{fig_high_dyn}
\end{figure}

\textbf{Burst-dynamics Generation.} Abrupt changes in scenes and motions lead to dynamics bursts, which include \textit{scene swaps} when the user changes applications and \textit{motion spikes} when the user experiences sudden acceleration or stops. To test the system's effectiveness in these scenarios, we generate two burst-dynamic scenarios that exhibit similar effects on the system. 
For scene swaps, we switch from the current scene to a new scene that contains over 2000 objects as shown in Fig.~\ref{fig_high_dyn}(A). We force rendering 10 frames of the new scene to mimic a surge in render demand, which is prevalent in abrupt scene-changing scenarios. 
For motion spikes, we connect the Zed Mini Stereo camera~\cite{zed_mini} and drop the camera from a fixed height, leading to a sudden acceleration depicted in Fig.~\ref{fig_high_dyn}(B). We increase the number of these burst dynamics during a 60-second runtime by injecting more occurrences randomly. We evaluate across all four baselines using two applications, Sponza and Gldemo on PC.

\begin{figure}[t]
\includegraphics[width=\linewidth]{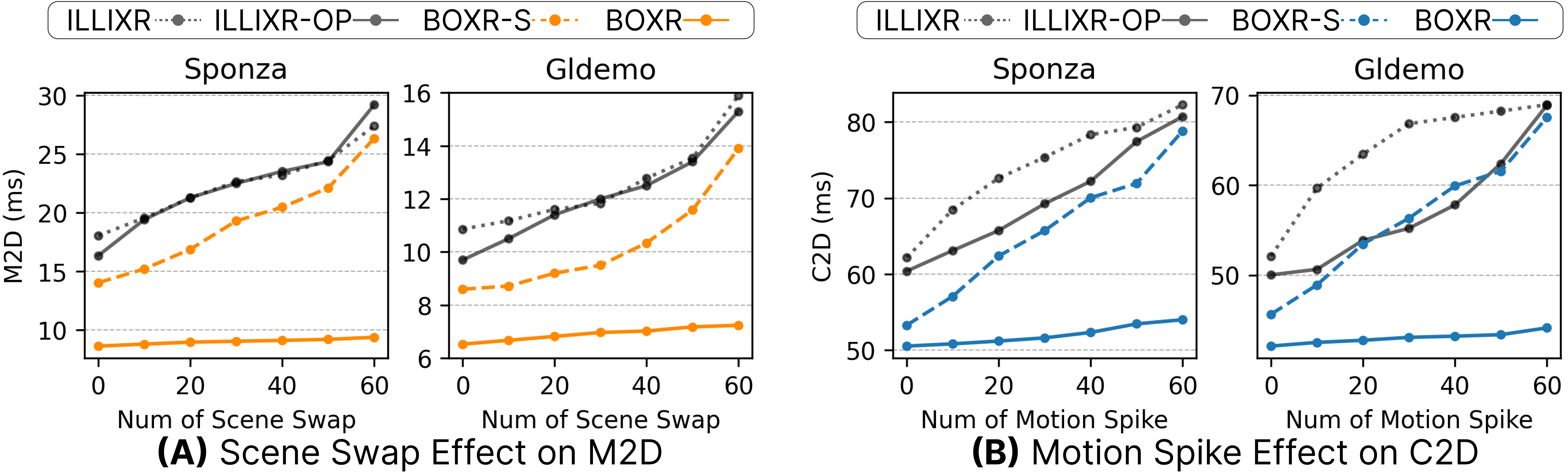}
\caption{Scene swap effect and motion spike effect}
\label{fig_eval_abrupt_motion}
\end{figure}

\textbf{Scene Swap Effect on M2D.} Scene swaps cause a significant render time increase, which substantially impacts the M2D metric illustrated in Fig.~\ref{fig_d_boxr_c1_c2}(A). We record the average M2D under an increased number of scene swaps during 60-second trials. As shown in Fig.~\ref{fig_eval_abrupt_motion}(A), the other three baselines experience up to a $57\%$ increase in M2D. In contrast, BOXR effectively limits the increase of M2D to at most $10\%$ when scene swaps occur every second. Compared to other baselines, BOXR keeps the render time below the designed budget through SFR, which effectively mitigates the rise in M2D even under abrupt scene swaps.

\textbf{Motion Spike Effect on C2D.} Motion spikes significantly increase the execution time of VIO, leading to a substantial rise in C2D depicted in Fig.~\ref{fig_d_boxr_c1_c2}(B). We, therefore, record the average C2D as the number of motion spikes gradually increases in Fig.~\ref{fig_eval_abrupt_motion}(B). The baselines show a significant increase in C2D, up to $48\%$, whereas BOXR only experiences a $6\%$ increase in C2D, owing to the use of MVIO, which effectively controls the execution time increase even during motion spikes. 

\begin{minipage}{0.92\linewidth}
\begin{shaded}
    \noindent \textbf{Robustness in extreme cases}: Stemming from Obs.~\ref{obs_3}, BOXR effectively limits the increase of M2D and C2D with its runtime adaptations in highly dynamic scenarios, proving its robustness within extreme use cases.
\end{shaded}
\end{minipage}

\subsection{Real-world Experiment}

\begin{figure}[t]
\includegraphics[width=\linewidth]{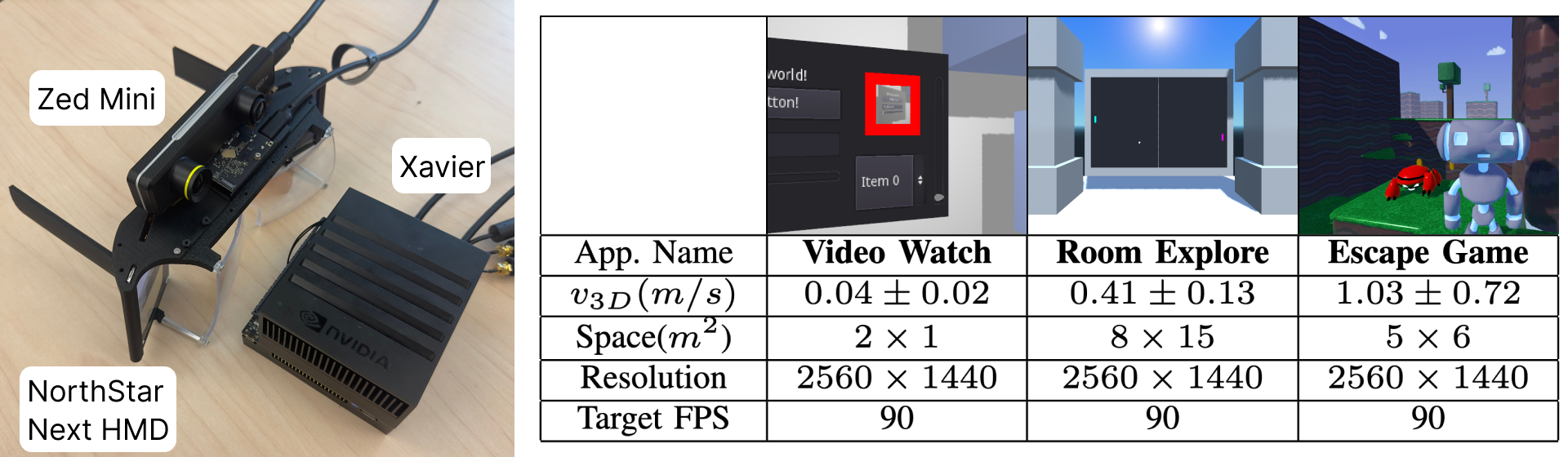}
\caption{Real-world Testcases setup}
\label{fig_real_world_setup}
\end{figure}

\textbf{Experiment Setting.} To closely match the state-of-the-art XR systems applications while maintaining freedom of movement, we connect the Zed Mini Camera~\cite{zed_mini} to the Xavier platform and design three XR applications using the Godot game engine~\cite{godot} to cover the different magnitudes of user motion, shown in Fig.~\ref{fig_real_world_setup}. Video Watch creates an interface that can play any static images or videos. Room Explore invites the users to roam freely in a 3D demo scene. Escape Game utilizes scenes from Platformer and makes the sprites chase and fire at the user, who needs to dodge and avoid being caught. We test the three applications on the default ILLIXR and the complete BOXR frameworks. We also adopt the 20ms M2D requirement and 80ms C2D requirement described in Sec.~\ref{subsec_characterizing_latency_metrics}, which indicate an unnoticeable head motion and body motion delay, respectively.

\begin{figure}[t]
\includegraphics[width=\linewidth]{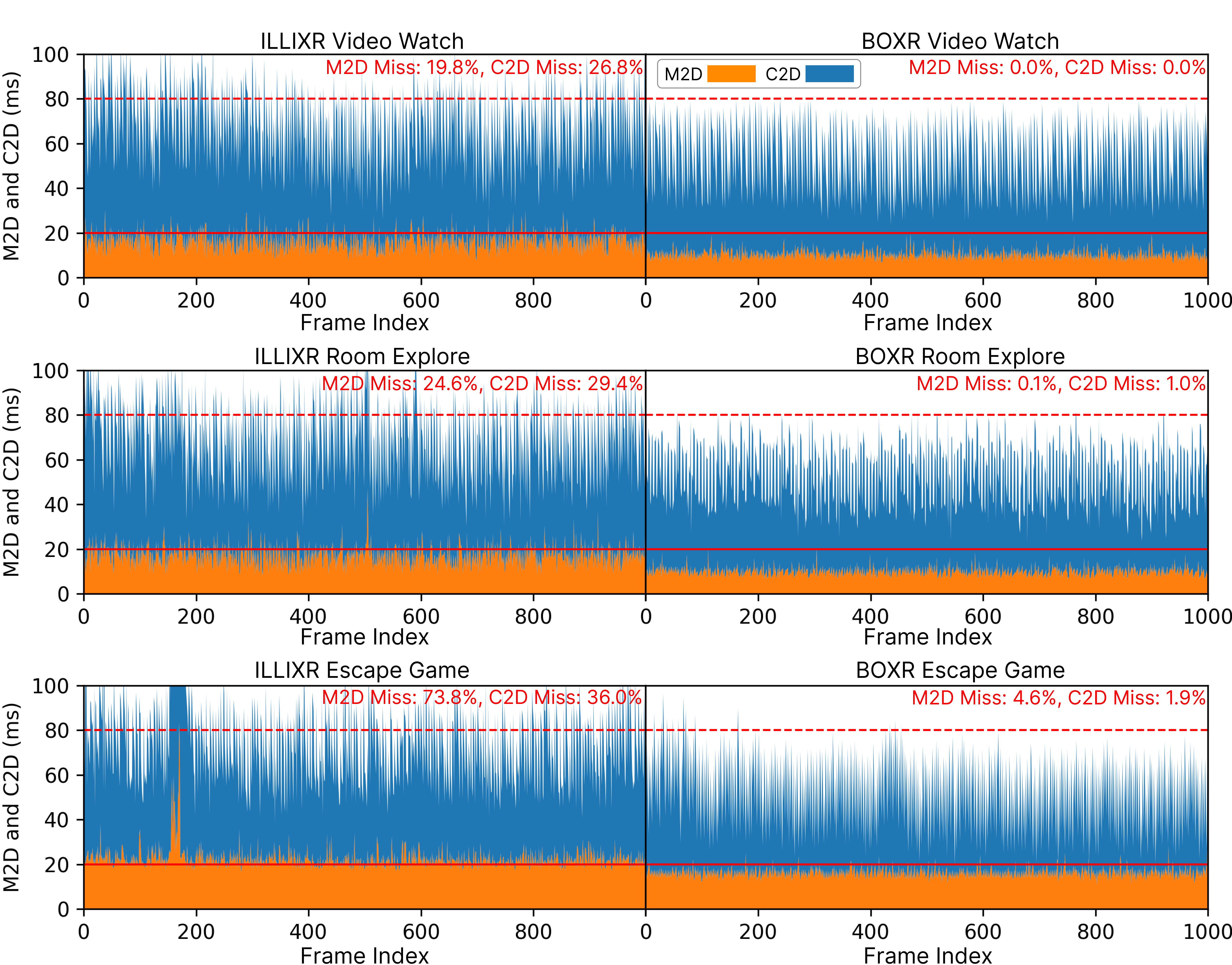}
\caption{M2D and C2D in Real-world Experiments}
\label{fig_eval_real_world}
\end{figure}

\textbf{M2D and C2D.} We sample M2D and C2D from 1000 frames during a 60-second trial and present the results in Fig.~\ref{fig_eval_real_world}. Across all applications tested, BOXR achieves up to a 42.6\% reduction in M2D and a 31.8\% reduction in C2D compared to ILLIXR. While no M2D and C2D in BOXR exceed their respective requirements in Video Watch, which involves low-magnitude motion, 4.6\% of M2D and 1.9\% of C2D values miss the requirements in Escape Game due to greater magnitude of motion. In extremely large motion around frame 200, ILLIXR experiences over 60ms of and 100ms of C2D, whereas BOXR's M2D stays below 20ms and C2D is maxed out at 90ms.

\begin{minipage}{0.92\linewidth}
\begin{shaded}
    \noindent \textbf{Applicability to real-world scenarios}: The consistently low M2D and C2D in real-world test cases prove the usability of BOXR in practical deployment.
\end{shaded}
\end{minipage}

\section{Related Work}

\textbf{Latency Metrics for XR}. Existing works have proposed multiple latency metrics in the context of distributed asynchronous systems similar to the pub-sub model used by XR systems~\cite{gunzel2021timing,kato2011timegraph}. Among these metrics, the most widely accepted metric for XR systems is M2D because of its direct association to motion-sickness~\cite{m2d_20_1,kundu2021study}. Existing work addresses this issue from various perspectives. Some employ reprojection methods for timely pose updates to reduce M2D~\cite{atw,min_m2d,hou2019head}. For example, \cite{atw} enables ATW on a commercial XR headset, while \cite{min_m2d} uses machine learning to predict and delay ATW execution. However, these methods assume fully preemptive GPU execution with no dependencies, which does not hold true in practical GPU systems~\cite{kato2011timegraph,wang2021balancing}. Dedicated hardware for tasks like ATW\cite{smit2009image,xie2019pim} increases power and complexity, challenging integration into existing XR frameworks. Offloading computation to nearby edge servers~\cite{ke2023collabvr,lai2017furion,meng2020coterie,liu2020firefly,xie2021q}, such as CollabVR~\cite{ke2023collabvr}, reduces M2D but requires high bandwidth and powerful PCs nearby. Despite these efforts, motion sickness persists due to neglect of C2D.

\textbf{XR Scene Rendering}.
The increased material resolution makes rendering every frame in raw resolution impractical~\cite{jabbireddy2022foveated}. Foveated rendering methods degrade peripheral viewport quality while maintaining central resolution~\cite{chaudhary2019ritnet, fujita2014foveated,weier2016foveated,turner2018phase,zheng2018perceptual,weier2018foveated,kim2019foveated,lee2020foveated}. RITnet~\cite{chaudhary2019ritnet} uses real-time semantic segmentation for eye-tracking and foveated rendering based on user gaze but lacks adaptive foveation based on scene dynamics. Reprojection-based methods run asynchronously to compensate for frame loss~\cite{atw,kijima2002reflex,mark1996post,peek2013more}, such as ~\cite{peek2013more} minimizing frame warping cost to improve frame rates. However, they often use previous frame data, leading to large C2D.

\section{Conclusion}
This paper presents BOXR: a Body and head motion Optimization framework for eXtended Reality. Building upon the three critical observations detailed in Sec.~\ref{sec_challenges}, BOXR employs a scheduling policy alongside two dynamic runtime adaptations in Sec.~\ref{subsec_mvio} and Sec.~\ref{subsec_sfr}. Through comprehensive comparison with ILLIXR, BOXR demonstrates significant performance improvement in controlled and real-world scenarios, proving the design's general effectiveness, robustness in extreme use cases, and adaptability to practical deployment. 


Despite all the benefits achieved by BOXR, there are still interesting directions for future work.
First, while BOXR focuses solely on software-level optimization, integrating software-hardware co-optimization could further enhance performance. Second, the benchmark we used primarily reflects data-intensive daily working scenarios but not GPU-intensive gaming scenarios. Addressing these will enable the development of a more comprehensive XR system.

\section*{Acknowledgement}
This work was supported by the National Science Foundation under Grants CPS 2230969, CNS 2300525, CNS 2343653, CNS 2312397, CNS 1943265.


\bibliography{ref}
\bibliographystyle{IEEEtran}

\end{document}